\documentclass[prb,reprint,amsmath,amssymb]{revtex4-2}

\usepackage{newtxtext}
\usepackage[varvw]{newtxmath}
\newcommand\smallblacktriangleright\blacktriangleright

\usepackage{graphicx}
\usepackage[caption=false]{subfig}

\usepackage[dvipsnames,svgnames,x11names,hyperref]{xcolor}
\usepackage{hyperref}
\hypersetup{
    colorlinks = true,
    linkcolor  = Blue,
    urlcolor   = Blue,
    citecolor  = Blue
}

\usepackage{bm,dsfont} 
\usepackage{physics,braket,siunitx,slashed}
\renewcommand\vec\bm

\newcommand{\vF}{v}

\newcommand{\subA}[2]{{#1}_{\scriptscriptstyle\textup{#2}}}
\newcommand{\subB}[2]{\overline{#1}_{\scriptscriptstyle\textup{#2}}}
\newcommand{\uS}{\subA{u}{S}}
\newcommand{\uA}{\subA{u}{A}}
\newcommand{\mS}{\subA{m}{S}}
\newcommand{\mA}{\subA{m}{A}}
\newcommand{\aS}{\subA{a}{S}}
\newcommand{\aA}{\subA{a}{A}}
\newcommand{\uSb}{\subB{u}{S}}
\newcommand{\uAb}{\subB{u}{A}}
\newcommand{\mSb}{\subB{m}{S}}
\newcommand{\mAb}{\subB{m}{A}}
\newcommand{\aSb}{\subB{a}{S}}
\newcommand{\aAb}{\subB{a}{A}}

\newcommand{\DeltaSinglet}{\Delta^{s\prime}_{\zeta}}
\newcommand{\DeltaDoublet}{\Delta^{s\prime\prime}_{\zeta}}

\begin{document}


\vspace*{0.9cm}
\title{Substrate-induced topological minibands in graphene}

\author{Tobias M. R. Wolf} 
\author{Oded Zilberberg} 
\author{Ivan Levkivskyi} 
\author{Gianni Blatter} 

\affiliation{%
    \mbox{Institute for Theoretical Physics, ETH Zurich, 8093 Zurich,
    Switzerland} \linebreak[4]
}%

\date{14 September 2018}

\begin{abstract}
The honeycomb lattice sets the basic arena for numerous ideas to implement electronic, photonic, or phononic topological bands in (meta-)materials. Novel opportunities to manipulate Dirac electrons in graphene through band engineering arise from superlattice potentials as induced by a substrate such as hexagonal boron-nitride. Making use of the general form of a weak substrate potential as dictated by symmetry, we analytically derive the low-energy minibands of the superstructure, including a characteristic 1.5 Dirac cone deriving from a three-band crossing at the Brillouin zone edge. Assuming a large supercell, we focus on a single Dirac cone (or valley) and find all possible arrangements of the low-energy electron and hole bands in a complete six-dimensional parameter space. We identify the various symmetry planes in parameter space inducing gap closures and find the sectors hosting topological minibands, including also complex band crossings that generate a valley Chern number atypically larger than one. Our map provides a starting point for the systematic design of topological bands by substrate engineering.
\end{abstract}

\maketitle
%


\section{Introduction}\label{sec:intro}
The hunt for materials with topological properties, originally rooted in 
two-dimensional quantum Hall systems \cite{Klitzing1986}, has been fueled by
numerous proposals for materials with electronic topological bands
\cite{kane2005,Hasan2010,Qi2011} and has recently sparked ideas for engineered
meta-materials hosting topological bands for electromagnetic
\cite{Haldane2008,Khanikaev2012,Ozawa2018} (photonic)  or elastic
\cite{Prodan2009,Susstrunk2016} (phononic) modes.  Many of these proposals are
based on the honeycomb lattice, which provides a natural host for topological
phenomena through various types of engineering, from the (dynamical) Haldane
model \cite{haldane1988,Oka2009}, recently realized both in photonic
\cite{Rechtsman2013} and cold atomic \cite{Jotzu2014} systems, to designer
dielectrics holding topological photons \cite{Wu2015}.  The topological
properties in these systems arise from band crossings or Dirac cones. In
time-reversal-symmetric systems, such cones appear in compensating pairs and
topological features cancel out. Nevertheless, topological properties manifest
in individual valleys or cones and are brought forward in the field of
valleytronics \cite{Rycerz2006,Xiao2007,Schaibley2016}---as with topological
materials, valleytronics can be engineered in non-electronic systems
\cite{Ni2018}.

In this paper, we investigate a generic valleytronic system where the Dirac electrons of the cone are engineered via a weak hexagonal substrate potential, with a well-known realization of such a system given by placing graphene on hexagonal boron nitride (G-hBN). Symmetry considerations on the substrate potential then define a six-dimensional pa- rameter space that describes all possible arrangements of minibands and their topological properties. Focusing on the six lowest electron and hole bands, the threefold symmetry of the scattering potential leads to a characteristic ``1.5'' Dirac cone deriving from three crossing bands as well as two strongly anisotropic two-band crossings. We discuss several pertinent examples for new topological band arrangements resulting from the atypical Berry curvatures generated by these crossings, including also situations with high valley Chern number.

While depositing graphene on a substrate can improve the electrical properties
of the film \cite{Geim2007,CastroNeto2009,Dean2010}, such simple manipulation
also allows for the (deliberate) tuning of its spectral properties. An example
that has received much attention recently is bilayer graphene with its flat
bands at ``magic twist-angles'' \cite{Li2010c,
SuarezMorell2010,Bistritzer2011,TramblyDeLaissardiere2012, Weckbecker2016},
domain-wall induced edge states \cite{Zhang2013,Yin2016,SanJose2014a,
Huang2018}, and superconducting properties \cite{Cao2018,Po2018,Roy2018}. In
this situation, the Dirac cones of both graphene layers contribute equally to
the physical phenomena. In contrast, when graphene is placed on an
\emph{insulating} substrate, e.g., silicon-carbide (SiC) or boron-nitride (BN),
the electronic states of the latter are separated far in energy. Consequently,
only the Dirac cones of graphene have to be considered, with corrections
induced by the substrate, such as a gap opening in the Dirac spectrum
\cite{Zhou2007,Giovannetti2007}. In particular, including the misfit between
the graphene lattice and the substrate lattice leads to a (``Moir\'e'')
superstructure that generates more complex reconstructions of the Dirac cones
into minibands \cite{Wallbank2013,SanJose2014a,Moon2014,Wallbank2015}: the
scattering of the graphene electrons on such a threefold symmetric and weak
substrate potential (e.g., G-hBN) naturally leads to the hybridization of
backfolded cones that results in secondary gap openings. The latter give birth
to conventional or even topological minibands
\cite{SanJose2014a,Song2015,Brown2017} and constitute the main focus of the
present work. The shape and character of these band arrangements depends
sensitively on the substrate-induced potential; here, rather than focusing on
specific lattice--substrate arrangements
\cite{SanJose2014a,Moon2014,Song2015,Brown2017}, we provide a complete map
relating the miniband structure with the scattering parameters of the substrate
potential under weak coupling conditions.

Similar multiband engineering has attracted interest in recent years, starting
with proposals to hybridize three (one Dirac cone plus a flat band)
\cite{Liu2011,Huang2011} and four bands (a double-Dirac cone)
\cite{Sakoda2012a} in photonic \cite{Wu2015} or phononic
\cite{Liu2012,Chen2014} metamaterials by exploiting properly tuned accidental
degeneracies. While these degenerate multiband configurations reside at the
$\Gamma$ point, our three-band mixing occurs near the $K$-points of a Dirac
material and involves three linear bands, corresponding to what we call a 1.5
Dirac cone, in allusion to the double-Dirac cone of
Refs.~[\onlinecite{Sakoda2012a, Wu2015,Liu2012,Chen2014}]. Alternatively, the
formation of what we call the 1.5 Dirac cones by the scattering of a Dirac
fermion on a hexagonal substrate potential can be understood in terms of the
formation of the new ``three-fermions'' of Ref.\ [\onlinecite{Bradlyn2016}].

\begin{figure}[t]
    \includegraphics[width=8.6cm]{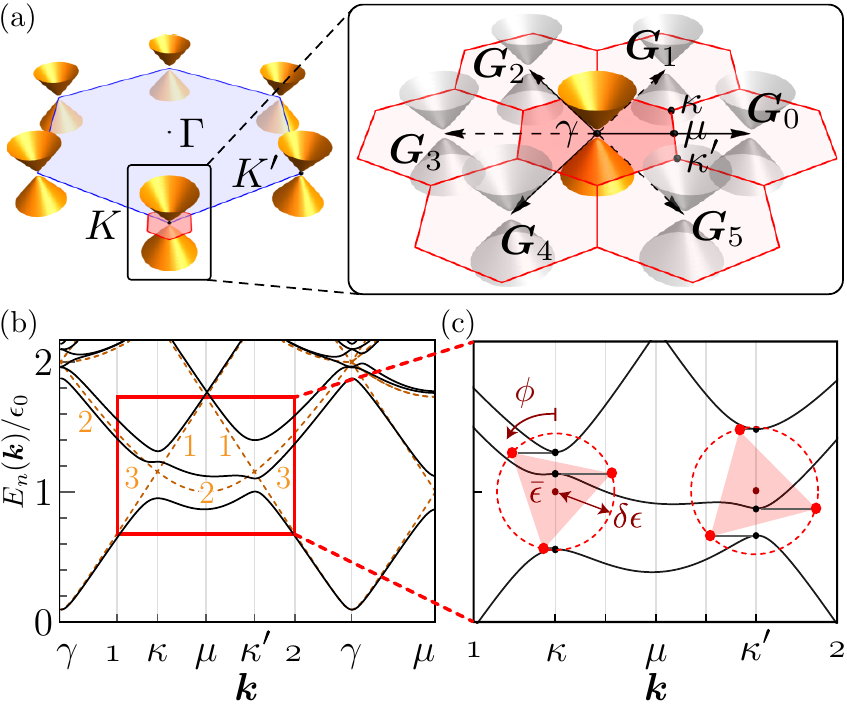}
	\caption{\label{fig:mBZ_hybridization2} 
    (a)~High- and low-energy description of Dirac-like particles subject to a
    substrate potential. The large Brillouin zone (BZ) of the host material
    (left) generates Dirac cones at the $K$- and $K'$-points (dark orange).
    Selecting one of the latter (here a $K$-point) defines the new
    $\gamma$ point of the mini-BZ (right) generated by the triangular substrate
    potential.  The scattering of the Dirac-like particle by the substrate
    produces a periodic mirroring (light gray cones) of the original Dirac cone
    and leads to band-hybridization at the boundary of the first mini-BZ (light
    shaded red). The high-symmetry points $\kappa$ and $\kappa'$ of the mini-BZ
    produce threefold degeneracies and the hybridization of the associated
    bands through the substrate generates bands with tunable topological
    properties. (b)~Electronic dispersion along the lines $\gamma \to \kappa
    \to \mu \to \kappa' \to \gamma \to \mu$, with dotted orange lines referring
    to pristine graphene (dispersion backfolded to the mini-BZ; numbers
    indicate degeneracies) and black solid lines showing the dispersion for a
    finite scattering potential $V(\vec{x})$ that lifts the degeneracies and
    opens gaps (we have chosen parameters $\Delta = 0.1\epsilon_0$ and $\mS =
    0.13 \epsilon_0,\,\mA = 0.03 \epsilon_0$; the red square delimits the
    region for the dispersions shown in Figs.\ \ref{fig:PhaseSpace_Symm} and
    \ref{fig:PhaseSpace_Asymm}). (c)~The band energies near the $\kappa$  and
    $\kappa'$ points appear as projections of three equidistant points (shaded
    red triangle) on a circle of radius $\delta\epsilon$ centered at
    $\overline{\epsilon}$ and rotated by $\phi$; see Eq.\
    \eqref{eq:Energies33_main}.
}
\end{figure}

Depositing graphene on a substrate with hexagonal symmetry generates both
(incommensurate) Moir\'e \cite{Xue2011,Decker2011} or (commensurate) grain-
boundary \cite{Yankowitz2012,Woods2014} superstructures. Below, we analyze how
such a superstructure splits an individual Dirac cone into minibands, see
Fig.\ \ref{fig:mBZ_hybridization2}(a). Exploiting that such a cone maps onto
itself under the combined action of inversion (I) and time reversal (T), we
can use symmetry arguments \cite{Wallbank2013} to characterize the scattering
potential. The latter then is described by six parameters that can be grouped
into two sets of three TI symmetric (TIS) and three TI antisymmetric (TIAS)
amplitudes, defining two three-dimensional parameter spaces.  The main goal of
the present work then is to find a systematic and complete map between the
six-dimensional parameter space of symmetry-allowed scattering potentials and
the ensuing miniband structure including its topological properties.

The three lowest electron and hole minibands derive from three backfolded
cones that mix at the $\kappa$ and $\kappa'$ points of the mini-Brillouin
zone, see Fig.\ \ref{fig:mBZ_hybridization2}(a).  A purely TI-symmetric
potential then splits the threefold degeneracies at the $\kappa$  and
$\kappa'$ points into combinations of a single cone and a parabolic band---the
mutual arrangement of the latter depends on the chosen parameters.  Turning on
a TI-antisymmetric component of the substrate potential leads to a splitting of
the remaining degeneracy of the cones and frees the Berry curvatures previously
hidden in the degeneracy points \cite{Nielsen1995}.  The Berry curvatures
deriving from the 1.5 cone sum up to values $\pm 1/4$ for the top and bottom
bands and averages to zero for the middle band, cf. numerical results in
Ref.~[\onlinecite{Song2015}], quite different from the usual weight $\pm 1/2$
characterizing a conventional Dirac-like cone. Our analytic derivation provides
deeper insights into the origin of these reduced contributions to the Chern
number. Finally, by proper tuning of parameters, we find values generating
electron or hole bands that are gapped away from other bands---appropriate
placement of the chemical potential within the minigap then allows for
realizing topological valley physics \cite{Xiao2007,Rycerz2006} with
minibands.  Furthermore, we find substrate configurations that generate such
isolated bands with a network of Berry curvature with a higher-than-one Chern
number.

In the following Sec. \ref{sec:dirac}, we set up our phenomenological model
Hamiltonian describing an isolated cone of Dirac-like particles subject to a
weak substrate potential with TI-symmetric $D_3$ and more general $C_3$
symmetries. We solve the problem analytically for the six low-energy electron 
and hole bands by folding back the neighboring unit cells in the Brillouin
zone; more exact band-structure calculations are done numerically with 62
bands, i.e., including higher-order reciprocal vectors. In Sec.\
\ref{sec:minibands}, we analyze the miniband geometries for the $D_3$ and $C_3$
symmetric potentials, emphasizing the geometric arrangements of the bands with
singlets and doublets at the $\kappa$ and $\kappa'$ points in the high symmetry
$D_3$ case and the Berry-curvature maps characteristic of the low-symmetry
($C_3$, TI-symmetry broken) situation. The latter derive from multiple band
crossings and we present an analytic calculation for the curvatures associated
with the various bands.  In Sec.\ \ref{sec:ins}, we present specific examples
where the substrate potential produces isolated minibands (with well defined
gaps separating bands) characterized by nontrivial Chern numbers. We summarize
our work and conclude in Sec.\ \ref{sec:concl}.

\section{Dirac-like particles in $C_3$ and $D_3$ symmetric potentials}
\label{sec:dirac}

We study an effective model describing the low-energy physics of Dirac
electrons subject to a weak hexagonal periodic potential. This situation is
realized by the triangular Moir\'e pattern resulting when graphene is deposited
on an insulating hexagonal substrate, such as boron-nitride.  To do so, we
consider a spinless Dirac-like particle described by a pseudospinor with
linear dispersion $T(\vec{k})$ moving in two dimensions in the presence of a
weak periodic potential $V(\vec{x})$,
\begin{align}
    \label{eq:ModelHamiltonian}
    H = \vF\,\hbar\vec{k}\cdot\vec{\sigma} + \Delta\,\sigma_3
    + \sum_{\ell=0,\dots,5} V_{\vec{G}_{\ell}}
    \,e^{i\vec{G}_{\ell}\cdot\vec{x}} ,
\end{align}
where $\sigma_j$ denote Pauli isospin matrices, $\vF$ is the Fermi velocity,
and we allow for a finite mass (or spectral gap) $\Delta$ which constitutes a
TIAS parameter.  We assume a smooth, threefold rotational-symmetric
potential with only one set of long-wavelength amplitudes $V_{\vec{G}_{\ell}}$
for the six reciprocal lattice vectors $\vec{G}_{\ell} = G \, \left[ \cos
\left({2\pi\ell} /{6}\right), \, \sin\left({2\pi\ell}/{6}\right) \right]$,
$\ell=0,\dots,5$. The reciprocal lattice constant $G = 4\pi/3L$, with $L$ the
real-space periodicity, defines an energy scale $\epsilon_0$ via the minimal
recoil momentum $G$ for elastic scattering,
\begin{align} \label{eq:recoil}
    \epsilon_0 \equiv \vF\,\hbar G/2.
\end{align}
We assume the latter  to be much larger than the mass $\Delta$ and the
amplitudes $V_{\vec{G}_{\ell}}$ of the potential. In building the Moir\'e pattern
in the G-hBN system, the (approximate) periodicity $L(\vartheta) = [(r-1)^2-4 r
\sin^2(\vartheta/2)]^{1/2} $ is determined by the ratio $r \gtrsim 1$ of the
lattice constants of graphene and the substrate, and their misfit angle
$\vartheta$ \cite{Wallbank2013,Moon2014}; note, that we deal with a slightly
incommensurate situation where the lattices do not match exactly on the
distance $L$. An exact match can be obtained in twisted bilayer graphene and
requires fine-tuning of the angle $\vartheta$ \cite{Moon2012}. In the
following, we ignore effects arising due to the quasi-periodicity appertaining
to a Moir\'e pattern.

The eigenmodes of the kinetic part $T(\vec{k}) + \Delta \sigma_3$ in the
Hamiltonian \eqref{eq:ModelHamiltonian} describe particles with dispersion
\begin{align} \label{eq:disp}
    \epsilon^{\pm}(\vec{k}) =\pm\sqrt{(\hbar \vF \,\vec{k})^2 + \Delta^2}
\end{align}
and associated momentum eigenstates
\begin{align}
    \label{eq:DiracEigenstates}
    \ket{\vec{k},\pm} = \frac{1}{\sqrt{2}}
    \begin{pmatrix}
        \phantom{\pm} \sqrt{1+\Delta/\epsilon^{\pm}(\vec{k})} \,
        e^{-i\,\varphi(\vec{k})/2} \\
        \pm \sqrt{1-\Delta/\epsilon^{\pm}(\vec{k})} \, e^{+
                i\,\varphi(\vec{k})/2}
    \end{pmatrix}
    \ket{\vec{k}},
\end{align}
where $\ket{\vec{k}}$ is a plane wave state with wave vector $\vec{k} =
(k_1,k_2)$. The phase $\varphi(\vec{k}) = \arg(k_1+i\,k_2)$ and signs $\pm$
refer to particle ($+$) and hole ($-$) bands, in the following specified by
the index $s = \pm 1$.

The most general expression for the small scattering amplitudes
$V_{\vec{G}_{\ell}}$ respecting threefold rotational symmetry takes the form
\cite{Wallbank2013}
\begin{align} \label{eq:Superpotential_FourerMatrix}
  V_{\vec{G}_{\ell}} = u_{\ell}+\,m_{\ell}\,\sigma_3+a_{\ell}\,
  (\vec{\hat{z}}\times i\,\vec{\hat{G}}_{\ell})\cdot \vec{\sigma},
\end{align}
where ${u_{\ell} = \uS + (-1)^\ell\,i\,\uA}$, ${m_{\ell} = \mA + (-1)^\ell\,
i\,\mS}$, and ${a_{\ell} = \aA + (-1)^\ell\,i\,\aS}$ define three complex
parameters with the hat $\vec{\hat{\cdot}}$ referring to normalized quantities
of unit amplitude. The real (imaginary) parts of the parameters define a
potential that is even (odd) under real space inversion. While $u_{\ell}$
quantifies the overall amplitude of the potential landscape, $m_{\ell}$
describes a periodic modulation of the Dirac mass $\Delta$. The parameters
$a_{\ell}$ are associated with a spatially periodic vector potential describing
the action of an out-of-plane pseudomagnetic field with the same spatial
periodicity as the (substrate) potential; in graphene, such a term can arise
due to nonuniform strain \cite{CastroNeto2009}. Note that local $U(1)$
symmetry in the Hamiltonian allows one to eliminate the longitudinal component
of this vector potential via a proper transformation of the wavefunction (see
Appendix \ref{sec:tr_gauge}); the parameters $a_{\ell}$ then describe the
transverse component of the vector potential after fixing the gauge.

The various components of the Hamiltonian can be grouped into two sets that
are defined through their transformation properties under the combined action
of time-reversal and spatial inversion, the TI-symmetric (TIS) parameters
$\uS,\mS,\aS$, and the TI-antisymmetric (TIAS) parameters $\Delta,\uA,\mA,\aA$
\cite{Wallbank2013}, the latter picking up a minus sign under the action of
TI. Dropping the TIAS parameters enhances the structural symmetry group from
$C_3$ to $D_3$. Note that T and I by themselves are not good symmetries of the
Hamiltonian \eqref{eq:ModelHamiltonian}, unless we include the host material's
second Dirac cone (e.g., at the time-reversed point $K'$, with the parameters
$(-\Delta,\uS,\uA,-\mA,-\mS,-\aA,-\aS)$ for the case of a time- reversal-
symmetric mass and potential).

Including the scattering by the potential $V(\vec{x})$, the free Dirac-like
spectrum is folded back in reciprocal space, defining the Brillouin zone (BZ)
shown on the right of Fig.\ \ref{fig:mBZ_hybridization2}(a).  The band
structure is obtained from diagonalizing the Bloch Hamiltonian
\begin{align}
    \label{eq:BlochHamiltonian}
    H(\vec{k}) \! = \! \sum_{i,j} \left[ \ket{\vec{k}_{ij}} T(\vec{k}_{ij})
        \bra{\vec{k}_{ij}}
        \! + \! \sum_{\ell=0}^5  \ket{\vec{k}_{ij} \! + \! \vec{G}_{\ell}}
        V_{\vec{G}_{\ell}} \bra{\vec{k}_{ij}} \right],
\end{align}
with $\vec{k}$ restricted to the first Brillouin zone and ${\vec{k}_{ij}
\equiv \vec{k} + i \vec{G}_0 + j \vec{G}_1}$, with $i,j$ integers, denoting the
original position in reciprocal space; see top right in Fig.\
\ref{fig:mBZ_hybridization2}.  Given a choice of scattering amplitudes, Eq.\
\eqref{eq:BlochHamiltonian} can be diagonalized numerically
\cite{Wallbank2013} including a sufficiently large set of bands $\{i,j\} \in
\mathbb{Z}^2$; see Fig.~\ref{fig:mBZ_hybridization2}(b). 

Alternatively, focusing on the lowest bands, useful insights can be gained
from an analytic solution involving only mixing of the three neighboring cells
sharing the $\kappa$ point and the $\kappa'$ point (equivalently, we denote
the latter by $\zeta \kappa$-points, $\zeta = \pm 1$, with $\kappa'$
equivalent to $-\kappa$). Including scattering induced by the potential
\eqref{eq:Superpotential_FourerMatrix} between the unperturbed states
$\ket{\zeta\vec{\kappa} + \vec{q},\pm}$, $\ket{\zeta(\vec{\kappa} -
\vec{G}_0)+\vec{q},\pm}$, and $\ket{\zeta(\vec{\kappa} -
\vec{G}_{1})+\vec{q},\pm}$, the many-band Bloch Hamiltonian
\eqref{eq:BlochHamiltonian} can be truncated to the lowest three electron
($s=1$) and hole ($s=-1$) bands described by
\begin{align}
    \label{eq:BlochHamiltonian_zeta}
    H^s_\zeta(\vec{q}) \! =
        \begin{pmatrix}
            \epsilon_{\zeta 0}^s & V_{\zeta 1}^s
            & {V_{\zeta 2}^{s}}^* \\
            {V_{\zeta 1}^{s}}^* & \epsilon_{\zeta 2}^s
            & V_{\zeta 0}^s \\
            V_{\zeta 2}^s & {V_{\zeta 0}^{s}}^* 
            & \epsilon_{\zeta 1}^s
        \end{pmatrix},
\end{align}
with the unperturbed energies $\epsilon_{\zeta j}^s \!=\! \epsilon^s(\zeta
\vec{\kappa}+\vec{q}_j)$ and matrix elements $V_{\zeta j}^s \!=\! \braket{\zeta
\vec{\kappa}+\vec{q}_j,s |V_{\vec{G}_0}| \zeta (\vec{\kappa}- \vec{G}_0) +
\vec{q}_j,s}$, where $\vec{q}_j = R_{2\pi j/3}\, \vec{q}$, $j=0,1,2$, are
$2\pi/3$-rotated $\vec{q}$-vectors.

Such a three-band degenerate perturbation theory provides a reliable
analytical solution near the BZ boundary, while the band structure is given by
the Dirac-like spectrum \eqref{eq:disp} near the $\gamma$ point.  By
diagonalizing \eqref{eq:BlochHamiltonian_zeta}, we find that the energies for
electrons and holes can be written in the form of projections of three points
on a circle of radius $\delta\epsilon_{\zeta s} (\vec{q})$ centered around the
mean energy $\bar{\epsilon}_{\zeta s}(\vec{q})$,
\begin{align}
    \label{eq:Energies33_main}
    \epsilon_{\zeta}^{s n}(\vec{q})
    = \bar{\epsilon}_{\zeta}^s(\vec{q}) +
    \delta\epsilon_{\zeta}^s(\vec{q})\;
    \cos[\phi_{\zeta}^s(\vec{q})+ 2\pi \iota^{sn}/3]
\end{align}
for the three bands $n = 1,2,3$ arranged in ascending order of excitation
energy; see Fig.\ 1(c). Here, the mean $\bar{\epsilon}_{\zeta}^s(\vec{q}) =
\sum_j \epsilon_{\zeta j}^s/3$ derives from the unperturbed energies averaged
over the $2\pi/3$-rotated $\vec{q}$ vectors, while the radius
$\delta\epsilon_{\zeta}^s (\vec{q}) = 2 (\sum_j[ 2\,(\Delta \epsilon^{s}_{\zeta
j})^2 + |V_{\zeta j}^s|^2]/3)^{1/2}$ involves the energy disbalance $\Delta
\epsilon_{\zeta j}^s = [\epsilon_{\zeta j}^s - \bar{\epsilon}_{\zeta}^s]/2$.
The offset angle $0\leq \phi_{\zeta}^s\leq \pi/3$ is given by
\begin{widetext}
\begin{align}\label{eq:Energies33_mainp}
    \phi_{\zeta}^s(\vec{q}) \! = \! \frac{1}{3} \!
    \cos^{-1}\left[
    \frac{\mathrm{Re}({\prod_j} V_{\zeta j}^s) + 4 \prod_j \Delta \epsilon_{\zeta j}^s  
    - \sum_j \Delta \epsilon_{\zeta j}^s\, \abs{V_{\zeta j}^s}^2}
    {[\delta\epsilon^{s}_{\zeta}(\vec{q})/2]^3} \right]
\end{align}
\end{widetext}
and the integer $\iota^{sn} = s(n-1/2) + 1/2$ ensures the proper band
ordering.  While the radius $\delta\epsilon_{\zeta}^s (\vec{q})$ defines the
magnitude of the splittings, the phase $\phi_{\zeta}^s(\vec{q})$ determines
their relative arrangements. The associated eigenfunctions can be found in a
closed analytic form as well; see Appendix \ref{sec:Eigensolutions}.

The three-band mixing described by \eqref{eq:BlochHamiltonian_zeta} determines
the structure of the minibands near the corresponding edge of the Brillouin
zone. In the absence of a scattering potential $V$, the three energies in Eq.\
\eqref{eq:Energies33_main} collapse to a triplet at $\zeta\kappa$ (i.e.,
$\vec{q}=\vec{0}$).  Deviations away from $\kappa$ are linear in $q$ and
locally define three planes that derive from the cutting of the three original
cones---these three planes define our 1.5 Dirac cone.  A finite scattering
potential $V$ lifts the threefold degeneracy near the $\zeta\kappa$ points; at
the high-symmetry points, the splitting derives from the Hamiltonian
\eqref{eq:BlochHamiltonian_zeta} at $\vec{q}=\vec{0}$,
\begin{align}
    \nonumber
    H^s_\zeta \! = \!
        \begin{pmatrix}
            \epsilon^s(\vec{\kappa}) & \frac23\DeltaSinglet\!+\!\frac{i}{\sqrt{3}}\DeltaDoublet  
            & \frac23\DeltaSinglet\!-\!\frac{i}{\sqrt{3}}\DeltaDoublet\\
            \frac23\DeltaSinglet\!-\!\frac{i}{\sqrt{3}}\DeltaDoublet& \epsilon^s(\vec{\kappa}) 
            & \frac23\DeltaSinglet\!+\!\frac{i}{\sqrt{3}}\DeltaDoublet\\
            \frac23\DeltaSinglet\!+\!\frac{i}{\sqrt{3}}\DeltaDoublet 
            & \frac23\DeltaSinglet\!-\!\frac{i}{\sqrt{3}}\DeltaDoublet & \epsilon^s(\vec{\kappa})
        \end{pmatrix} ,
\end{align}
with  
\begin{align}
    \label{eq:w_approx}
    \DeltaSinglet=3/2\,\mathrm{Re}  V_{\zeta}^s &= \frac{3}{4}
    (-\uS + \zeta\, \sqrt{3} \mS + \zeta \, s \, 2\aS),
    \nonumber                                                                \\
    \DeltaDoublet=\sqrt{3}\, \mathrm{Im}  V_{\zeta}^s &= \frac{\sqrt{3}}{2}
    (\zeta\, \uA + \sqrt{3}\mA + s \, 2\aA)
\end{align}
derived from the scattering amplitudes $V_{\zeta}^s(\vec{q}=\vec{0})$ (we
assume a vanishing Dirac mass $\Delta=0$, see Sec. \ref{sec:Delta} for
results with a finite $\Delta$).  Diagonalizing $H^s_\zeta$, we find the
energy splittings
\begin{align}
    \label{eq:Energies33_main0}
    \epsilon_{\zeta}^{sn}|_\mathrm{singlet}
     & = \epsilon^{s}(\vec{\kappa}) + \frac{2}{3} \, \DeltaSinglet,
\nonumber      \\
    \epsilon_{\zeta}^{sn}|_\mathrm{doublet}
     & = \epsilon^{s}(\vec{\kappa}) -  \frac{1}{3}\, \DeltaSinglet \pm
\frac{1}{2}
    \, \DeltaDoublet,
\end{align}
where the splittings $\DeltaSinglet$ and $\DeltaDoublet$ are associated with
the TIS parameters $\uS$, $\mS$, and $\aS$ and the TIAS parameters $\uA$,
$\mA$, and $\aA$, respectively. 

In a TI-symmetric situation, we have $\DeltaDoublet=0$ and the original
triplet splits into a singlet and a doublet separated by $\DeltaSinglet$. If
$\DeltaSinglet>0$ ($<0$), the singlet will be higher (lower) in energy than
the doublet, which is equivalent to an offset angle $\phi_\zeta^s(\vec{0})=0$
($\pi/3$); in the following, we will denote these arrangements by
$\blacktriangle$ ($\blacktriangledown$), corresponding to the red shaded
triangles in Fig.\ \ref{fig:mBZ_hybridization2}(c). In the opposite case with
only finite TIAS parameters, we find that $\DeltaSinglet=0$ and the triplet
fully splits into singlets in a symmetric fashion. This splitting is
controlled by $\DeltaDoublet$ and comes with the offset phase
$\phi_\zeta^s(\vec{0})=\pi/6$ and thus will be denoted with the symbol
$\smallblacktriangleright$. A general TI-symmetry broken case will involve all
parameters and leads to an interplay between the singlet-doublet splitting
$\DeltaSinglet$ and the doublet splitting $\DeltaDoublet$. In such a
situation, the angle $\phi_\zeta^s(\vec{0})$ can assume any value. We
summarize the above discussion in Table~\ref{tab:TIsymmetry}.
\begin{table}[!b]
    \caption{\label{tab:TIsymmetry}
    Parameters driving the energy splittings $\DeltaSinglet$ and
    $\DeltaDoublet$ at the $\zeta\kappa$ points.  The symbols
    $\blacktriangle$, $\blacktriangledown$, $\smallblacktriangleright$ indicate the
    arrangement of split energies, doublet below singlet when $\DeltaSinglet
    >0$, doublet above singlet for $\DeltaSinglet <0$, and full symmetric
    splitting for $\DeltaSinglet = 0$ and $\DeltaDoublet \neq 0$,
    respectively. These arrangements are dictated by the offset angle
    $\phi_\zeta^s(\vec{0})$ [Eq.~\eqref{eq:Energies33_mainp}].
    }
    \centering
    \begin{ruledtabular}
        \begin{tabular}{r|l|l|l|l}
                 & Parameters  & $\DeltaSinglet$ & $\DeltaDoublet$ & $\phi_{\zeta}^s(\vec{0})$ \\
        \hline
        TIS    & $\uS,\mS,\aS$ & $>0, <0$ & $0$ 
                 & $0$ ($\blacktriangle$), $\pi/3$ ($\blacktriangledown$) \\
        TIAS   & $\uA,\mA,\aA,\Delta$ & $0$ & $\neq 0$ & $\pi/6$ ($\smallblacktriangleright$)
        \end{tabular}
    \end{ruledtabular}
\end{table}

A similar analysis can be done at the $\mu$  point, where it is sufficient to
consider two bands only; the Hamiltonian mixing the corresponding states
$\ket{\vec{\mu},\pm}$ and $\ket{-\vec{\mu},\pm}$ takes the form
\begin{align}
    \label{eq:BlochHamiltonian_mu}
   H^s_\mu \! \approx \! \begin{pmatrix}
   \epsilon^s(\vec{\mu}) & \!\!\!\!\frac12 \bar\Delta^s  \\
   \frac12{{\bar\Delta}}^{s *} \;\;
   & \!\!\!\! \epsilon^s(\vec{\mu})
\end{pmatrix},
\end{align}
where $\bar{\Delta}^s$ is the band gap at $\mu$ induced by the potential, 
\begin{align}
   \bar{\Delta}^s\!=\! 2 s (\aS\!-\!i\,\aA)\! -\! 2 (\mS\!-\!i\,\mA).
\end{align}
Diagonalizing \eqref{eq:BlochHamiltonian_mu}, we obtain the energy splitting
at the $\mu$  point in the form
\begin{align}
    \label{eq:Energies22}
    \epsilon_{\mu}^{sn}|_\mathrm{doublet}
     & = \epsilon^{s}(\vec{\mu}) \pm \frac12\abs{{\bar{\Delta}}_s}.
\end{align}


\section{Bloch bands along $\kappa$--$\mu$--$\kappa'$}\label{sec:minibands}

As illustrated in Figs.~\ref{fig:PhaseSpace_Symm} and
\ref{fig:PhaseSpace_Asymm}, the TI symmetric parameters split the triplet at
$\kappa$ and $\kappa'$ into a doublet (with the doublet degeneracy protected by
the TI symmetry) and an additional singlet, while the antisymmetric terms in
$V$ additionally split the remaining doublet---it is the latter splitting that
generates the topological properties of the minibands by breaking the TI
symmetry.

\subsection{$D_3$ symmetry}\label{sec:IS}

\begin{figure}[t]
    \includegraphics[width=8.6cm]{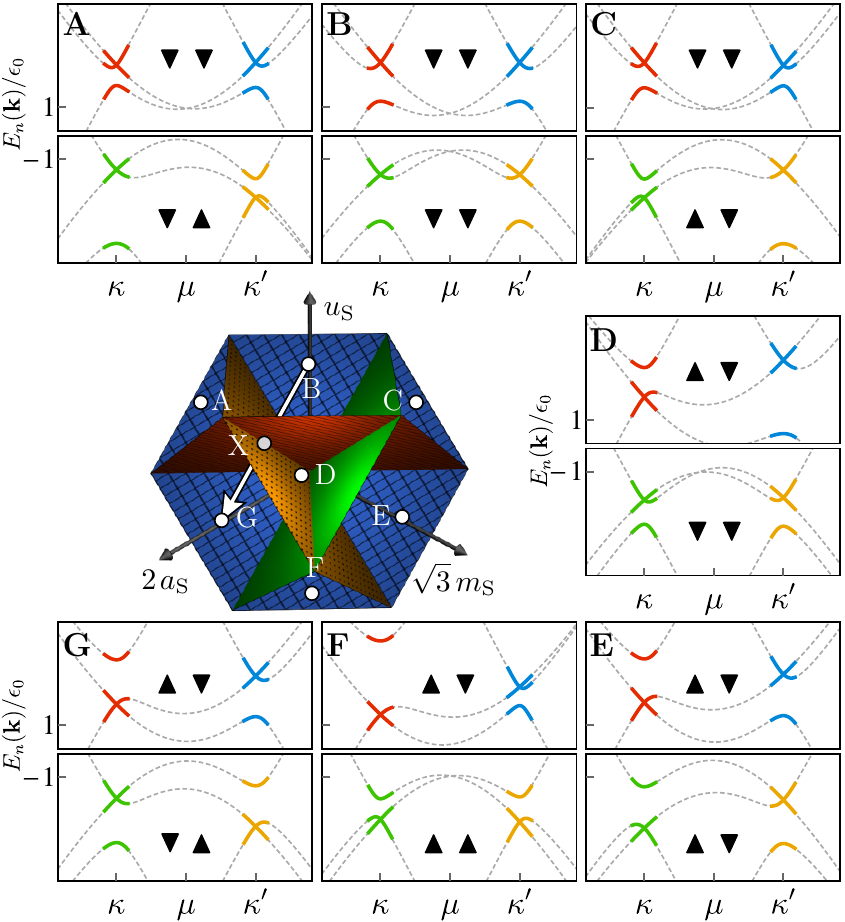} 
    \caption{\label{fig:PhaseSpace_Symm}
    Singlet--doublet gaps at the $\kappa$ and $\kappa'$ points depending on
    the TIS parameters $\uS,\mS,\aS$ with vanishing $\Delta,\uA,\mA,\aA$.
    The doublets locally define (warped) cones around the $\kappa$ and
    $\kappa'$ points in the 2D BZ, while the singlet is the projection of a
    (warped) paraboloid; see Fig.\ \ref{fig:TriangleRotations}(b). The
    relative band arrangement is indicated as $\blacktriangledown$
    ($\blacktriangle$) when the doublet is higher (lower) in energy than the
    singlet.  The four planes in the central figure mark the points of
    threefold degeneracy at the $\kappa$  and $\kappa'$ points where gap
    closures allow for band rearrangements. The surrounding insets show the
    different types of band splittings along the line $1 \to \kappa \to \mu
    \to \kappa' \to 2$ in the relevant energy window of the BZ [see Fig.\
    \ref{fig:mBZ_hybridization2}] that occur for parameters away from the
    planes (dispersions are calculated numerically accounting for 64 Bloch
    bands). The color code for the bands is red for $(\zeta,s) = (+,+)$, blue
    for $(\zeta,s) = (-,+)$, green for $(\zeta,s) = (+,-)$, and orange for
    $(\zeta,s) = (-,-)$.  Crossing a plane in parameter space inverts the
    bands ($\blacktriangledown\leftrightarrow\blacktriangle$) of the
    respective color in the band structure, e.g., going from {\bfseries A} to
    {\bfseries G} the crossing of the red plane rearranges the bands in
    $(\zeta,s) = (+,+)$. The evolution of the bandstructure
    along the line {\bfseries B} to {\bfseries G} in parameter space is shown
    in Fig.\ \ref{fig:EvolutionPlaneCrossing}.
    }
\end{figure}

We first focus on the TI-preserving situation with $\uA = \mA = \aA = \Delta =
0$.  The remaining TIS parameters define the 3D parameter space
$(\uS,\mS,\aS)$ shown in Fig.\ \ref{fig:PhaseSpace_Symm}. The four planes mark
parameters for which residual triplet degeneracies remain at $\kappa$ and
$\kappa'$, i.e., they signal singlet--doublet gap closures ($\DeltaSinglet =
0$).  These planes\footnote{Including higher bands leads to a deformation of
the planes.} compartmentalize the three-dimensional parameter space into $n =
14$ regions with different characteristic band arrangements, seven of which are
shown in the maps {\bf A} to {\bf G} in Fig.\ \ref{fig:PhaseSpace_Symm}.
Each of these maps is characterized by a different arrangement of singlets and
doublets in the four $(\zeta,s)$ sectors ($\kappa$ versus $\kappa'$, electrons
versus holes), with the singlet either above the doublet ($\DeltaSinglet>0$,
$\blacktriangle$) or vice versa ($\DeltaSinglet<0$, $\blacktriangledown$).
The maps {\bfseries A} to {\bfseries G} show the situation with an arrangement
$\blacktriangledown$ for positive energies at $\kappa'$ 
($\Delta^{+\,\prime}_{-} < 0$); the configurations with a $\blacktriangle$
instead appear for $\Delta^{+\,\prime}_{-}\sim -(\uS + 2\aS + \sqrt{3} \mS) >
0$. Note that the configurations $(\blacktriangledown
\blacktriangledown;\blacktriangle \blacktriangle)$ and $(\blacktriangle
\blacktriangle; \blacktriangledown \blacktriangledown)$ do not occur, since
all bounding planes are meeting in the origin of the parameter space.

\begin{figure}[b]
    \includegraphics[page=1,width=8.6cm]{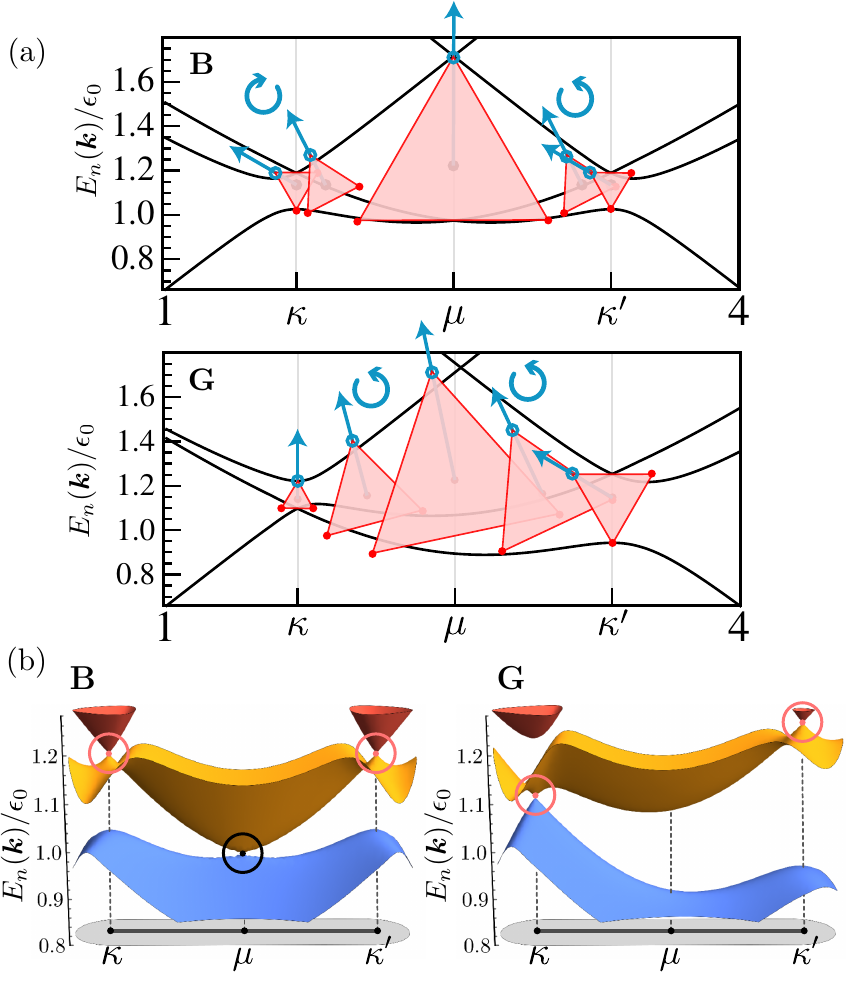}
    \caption{\label{fig:TriangleRotations}
    Band structure for the lowest three conduction bands ($s=+$) near the BZ
    edge ($\kappa$ -$\mu$  -$\kappa'$) with TIS-symmetric parameters
    corresponding to the maps {\bfseries B} and {\bfseries G} in
    Fig.\ \ref{fig:PhaseSpace_Symm}.
    (a) Band energies deriving from projections of three points on the circle
    \eqref{eq:Energies33_main}; the corresponding triangle (light red shading)
    rotates with the offset angle $\phi_\zeta^s(\vec{0})$, see Eq.\
    \eqref{eq:Energies33_mainp}, along the path $\kappa\to\mu\to\kappa'$ in
    reciprocal space. Two inequivalent situations are shown, with the same
    arrangement $\blacktriangledown\blacktriangledown$ between $\kappa$ and
    $\kappa'$ as featured in {\bfseries B} and with an opposite arrangement
    $\blacktriangle\blacktriangledown$ as in {\bfseries G}. In the former
    case, the triangle undergoes a rotation from $\pi/3\to0\to\pi/3$ along
    $\kappa\to\mu\to\kappa'$, while in the latter case it evolves
    unidirectionally $0\to\pi/3$.
    (b) 3D illustration of the band structure near the BZ edge
    showing the nontrivial geometries of the band touchings
    at $\kappa$ and $\kappa'$ (light red open circles) and at $\mu$ (black
    open circle) for parameters as in the maps {\bfseries B} and {\bfseries G}
    of Fig.\ \ref{fig:PhaseSpace_Symm}.
    }
\end{figure}

For a given point in the TIS parameter space, the $\blacktriangledown$ and
$\blacktriangle$ configurations at $\kappa$ and $\kappa'$ are smoothly related
through the evolution of the angle $\phi^s$ (see Fig.\
\ref{fig:TriangleRotations}). Configurations with the same singlet--doublet
arrangement at $\kappa$ and $\kappa'$ ($\blacktriangledown\blacktriangledown$
or $\blacktriangle\blacktriangle$) as in case {\bfseries B} have an intermediate level
crossing with the phase $\phi^s$ continuously changing either from
${\phi_{+}^s(\vec{0})=0}$ to $\pi/3$ and back to ${\phi_{-}^s(\vec{0})=0}$ or
from ${\phi_{+}^s(\vec{0})=\pi/3}$ to $0$ and back to
${\phi_{-}^s(\vec{0})=\pi/3}$.  Configurations that change the
singlet--doublet arrangement as in case {\bfseries G} ($\blacktriangledown\blacktriangle$
or $\blacktriangle\blacktriangledown$) have no intermediate level crossing and
the phase evolves unidirectionally from ${\phi_{+}^s(\vec{0})=0}$ to
${\phi_{-}^s(\vec{0})=\pi/3}$ or from ${\phi_{+}^s(\vec{0})=\pi/3}$ to
${\phi_{-}^s(\vec{0})=0}$.

\begin{figure}[b]
    \includegraphics[page=2,width=8.6cm]{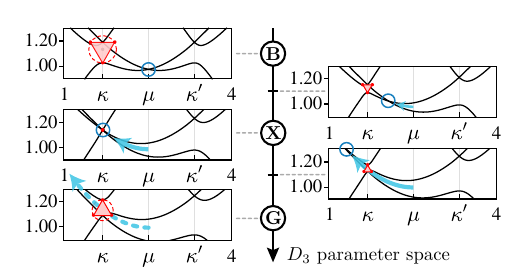}
    \caption{\label{fig:EvolutionPlaneCrossing}
    Evolution of the lowest three conduction bands ($s=+$) near the BZ edge
    ($\kappa$ -$\mu$  -$\kappa'$) when going from the TIS-symmetric parameter
    map {\bfseries B} to {\bfseries G} via a straight line as shown in
    Fig.~\ref{fig:PhaseSpace_Symm}.
    The singlet--doublet splitting at $\kappa$ (inverted red triangle) shrinks
    to zero upon approaching the point {\bfseries X} and opens up in an
    inverted geometry upon continuing towards {\bfseries G} (red triangle). At
    the same time, the band crossing at $\mu$ in {\bfseries B} (blue open
    circle) moves towards the $\kappa$ point (see cyan arrows). Passing
    through the triple degeneracy at {\bfseries X}, the band crossing moves
    further to higher bands as the parameters approach {\bfseries G}.
    }
\end{figure}

Tuning the TIS parameters across the plane associated with a given $\zeta$,
$s$ inverts the corresponding singlet--doublet configuration
($\blacktriangledown \leftrightarrow \blacktriangle$) by going through a gap
closing and reopening. As the gap vanishes at $\DeltaSinglet = 0$, the radius
$\delta\epsilon_{\zeta}^s(\vec{0})$ of the circle in
\eqref{eq:Energies33_main} goes through zero and the offset phase
$\phi_{\zeta}^s(\vec{0})$ flips by $\pi/3$. An example for for such a gap
closure and reopening when going from the maps {\bfseries B} to  {\bfseries G}
in Fig.\ \ref{fig:PhaseSpace_Symm} is shown in Fig.\
\ref{fig:EvolutionPlaneCrossing}.  Note that the configuration
$\blacktriangledown\blacktriangledown$ of {\bfseries B} features an
intermediate band crossing along $\kappa$ --$\mu$  --$\kappa'$ while the
configuration to $\blacktriangle\blacktriangledown$ in {\bfseries G} does not.
Hence, while moving across the plane in parameter space, this intermediate
crossing must continuously move into a higher band when passing through the
triple degeneracy.

\subsection{$C_3$ symmetry}\label{sec:IAS}

\begin{figure}[t]
    \includegraphics[width=8.6cm]{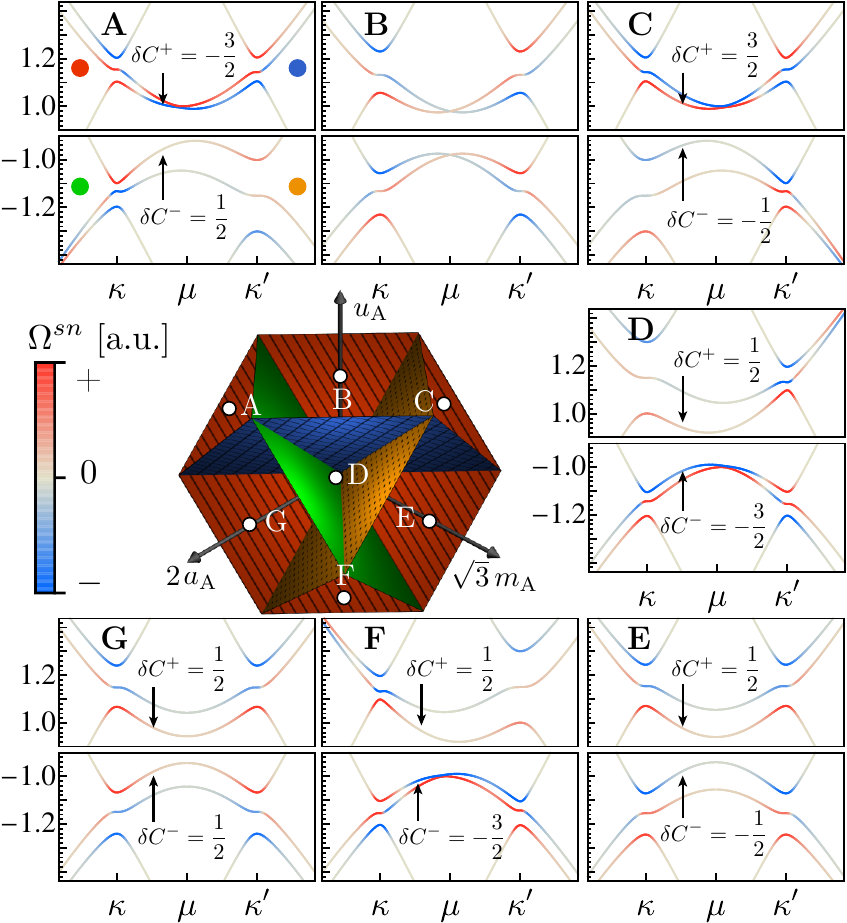} 
    \caption{\label{fig:PhaseSpace_Asymm}
    Doublet splittings at the $\kappa$ and $\kappa'$ points depending on the
    parameters $\uA,\mA,\aA$ for the massless TI-antisymmetric situation and
    with vanishing parameters $\Delta,\uS,\mS,\aS$. Colored segments describe
    the local Berry curvature $\Omega^{sn}(\vec{k})$ [see Sec. 
    \ref{sec:Chern}] of the $n$-th electron  and hole bands, with red (blue)
    denoting positive (negative) values, see $\Omega^{sn}$ color scale.  The
    four planes in the central figure mark the points of threefold degeneracy
    at the $\kappa$  and $\kappa'$ points where gap closures allow for band
    rearrangements and exchange of Berry curvature (see color-code in
    \textbf{A} relating the four $(\zeta,s)$-sectors with the relevant planes
    for gap closure).  The same color code as in Fig.\
    \ref{fig:PhaseSpace_Symm} has been chosen, e.g., blue for $(\zeta,s) =
    (-,+)$, implying a corresponding exchange of the Berry curvature when going
    from \textbf{B} to \textbf{D}. The fractions $\delta C^s$ denote the
    integrated Berry curvatures for the lowest ($n=1$) electron  and hole bands
    arising from the vicinity of the $\kappa$ -$\mu$  -$\kappa'$ points; adding
    the contribution $\delta C_{\gamma}^s =\pm 1/2$ from the vicinity of the
    $\gamma$ point (not shown) provides the Chern number $C^s= \delta
    C_{\gamma}^s + \delta C^s$ for the lowest bands. In \textbf{B}, the bands
    cross at $\mu$ and the Abelian Berry curvature is not well-defined. 
    }
\end{figure}

Next, we address the TI-antisymmetric case, i.e., with finite TIAS
parameters $\uA,\mA,\aA$ and vanishing TIS parameters $\uS = \mS = \aS = 0$
as well as $\Delta = 0$. This defines a second 3D space of antisymmetric
parameters $\uA,\mA,\aA$ describing spectra at $\kappa$  and $\kappa'$ points,
where the singlet in Eq.\ \eqref{eq:Energies33_main0}  remains unchanged while
the doublet is split symmetrically away; see Fig.\ \ref{fig:PhaseSpace_Asymm}.
Most interestingly, this gap opening induces finite Berry curvatures (see
Sec. \ref{sec:Chern} for a detailed analysis) in the 1.5 cone and generates
the curvature maps {\bfseries A} to {\bfseries G} shown in Fig.\
\ref{fig:PhaseSpace_Asymm}.  Shown are the configurations with equivalent
Berry curvatures in the $(\zeta,s) = (+,+)$ sector, i.e., for
$\Delta^{+\,\prime\prime}_{+} \sim (\uA + 2\aA + \sqrt{3} \mA) > 0$;
configurations with reverse Berry curvatures are realized for negative values
$\Delta^{+\,\prime\prime}_{+} < 0$.  Again, the four planes crossing at the
origin define the locations of triple-degenerate bands at the $\kappa$  and
$\kappa'$ points where bands rearrange when $\DeltaDoublet=0$.  As before, the
evolution of the spectra when moving between $\kappa$ and $\kappa'$ points and
when changing the antisymmetric parameters across one of the four
triplet-degenerate planes can be understood in terms of the rotation and
inflation/deflation of the circle defining the energies in
\eqref{eq:Energies33_main}.

In general, we deal with the situation that is neither purely symmetric nor
purely antisymmetric with respect to TI symmetry and thus we have to cope with all six
TIS and TIAS parameters assuming finite values.  We then have to consider
the interplay of the two cases. As we have seen, the TIS terms give rise to
doublet--singlet gaps, while the TIAS terms control the splitting of the doublet.
The magnitude of both splittings is determined by the distance of the
configuration point to a particular plane in the respective parameter space.
Knowing the two points in the parameter spaces of
Fig.~\ref{fig:PhaseSpace_Symm} and Fig.~\ref{fig:PhaseSpace_Asymm} allows one
to quickly determine the band configurations at the $\kappa$  and
$\kappa'$ points as well as their associated Berry curvatures.

\subsection{Finite mass $\Delta$}\label{sec:Delta}
In the above discussions of $D_3$- and the $C_3$-symmetric cases, we have
assumed a vanishing Dirac mass $\Delta$. Relaxing this assumption to a
situation where $\Delta$ is finite and of similar or smaller magnitude as the
other $3+3$ parameters, we find two effects: first, a gap opening at the
$\gamma$ point, freeing additional Berry curvature and allowing one to define an
(Abelian) Berry curvature for each band (see Sec. \ref{sec:Chern}). The
periodic potential can perturbatively modify this mass gap, $\Delta \to
\bar{\Delta}\approx\Delta+\Delta^{(3)}$; as the correction appears only in the
third order of the scattering parameters,
\begin{align}\label{eq:D3}
  \Delta^{(3)} &= -3\, \mA\frac{\mA^2-3
   \mS^2-\left(\uS^2-\uA^2\right)+4
   \left(\aS^2-\aA^2\right)}{\epsilon_0^2} \nonumber \\
   &\quad -3\, \mS\frac{
   2 \uS \uA+8 \aS \aA}{\epsilon_0^2},
\end{align}
the renormalization of the mass gap at $\gamma$ is small and can usually be
neglected. Second, a finite mass $\Delta$ produces a (again small) correction
in the band arrangement at the Brillouin zone edge.  Interestingly, the previous
classification of Figs.\ \ref{fig:PhaseSpace_Symm} and
\ref{fig:PhaseSpace_Asymm} remains quantitatively correct under proper
replacement of the bare TIS and TIAS parameters with their (slightly)
renormalized counterparts
\begin{align}
    \nonumber
    \uSb & = \uS+\alpha\,\mA, \quad
    \uAb =\uA+\alpha\,\mS,
    \\  \label{eq:ren_par}
    \mAb & = \mA+\alpha\,\uS, \quad
    \mSb =\mS+\alpha\,\uA
    \\ \nonumber
    \aAb & = \aA \sqrt{1-\alpha^2}, \quad
    \aSb = \aS \sqrt{1-\alpha^2},
\end{align}
where $\alpha \equiv \Delta / \epsilon^+(\vec{\kappa}) \ll 1$. This result can
be obtained by evaluating Eq.~\eqref{eq:w_approx} with a finite but small
$\Delta$. It should be noted, however, that in such a case the $D_3$-symmetry
for states near the BZ boundary holds only approximately: while the
doublet is no longer symmetry protected, the $\Delta$-induced splitting near
the BZ edge will be at least an order of magnitude smaller than any other
discussed gap contribution. 

\subsection{Berry curvatures and Chern numbers}\label{sec:Chern}
The breaking of either inversion or time-reversal symmetry in our single-valley
model \eqref{eq:ModelHamiltonian} allows for a finite Abelian 2D Berry
curvature
\cite{Xiao2010}
\begin{align}
\label{eq:BerryCurvature2D}
   \Omega^{sn}(\vec{k}) 
    &= i \!\!\! \sum_{s'n'\neq sn}\!\!\!\!\frac{\braket{sn|\partial_1 H|s'n'}
    \braket{s'n'|\partial_2 H|sn}-\left(1\!\leftrightarrow\! 2\right)}
    {[\epsilon^{sn}-\epsilon^{s'n'}]^2},
\end{align}
where $sn$ is the band index with associated energy $\epsilon^{sn}(\vec{k})$
and wave function $\ket{\vec{k},sn}$ and $\partial_{1,2}$ denote derivatives
along $k_1$ and $k_2$; we have suppressed the variable $\vec{k}$ in the right-
hand side of Eq.\ \eqref{eq:BerryCurvature2D}.  When bands do not cross, their
Berry curvature is well-defined and we can assign a Chern number
\begin{align}
\label{eq:ChernNumber}
   C^{sn} = \frac{1}{2\pi} \int_{\textrm{BZ}}\dd[2]{k} 
   \Omega^{sn}(\vec{k}) \in \mathbb{N}
\end{align}
to each separate band $sn$. In the following, we determine the local Berry
curvatures near the high-symmetry points $\gamma$, $\kappa$, $\mu$, and
$\kappa'$ in reciprocal space and integrate their contributions to the Chern
number.  For that matter, the contribution for the lowest electron  and hole
bands at the $\gamma$ point involves a standard calculation~\cite{Xiao2010};
the result is dominated by the Dirac mass $\Delta$ and results in a local Berry
curvature
\begin{align} \label{eq:Om-gamma}
    \Omega^{s}_{\gamma}(q) = \frac s2 \frac{(\hbar v)^2 \, \Delta}
    {[\Delta^2 + (\hbar v)^2 \, q^2]^{3/2}},
\end{align}
where the wavevector $q$ denotes the deviation from the $\gamma$ point. The
result \eqref{eq:Om-gamma} describes a smeared (by $\Delta/\hbar v$)
$\delta$-function in 2D with weight 1/2 and hence contributes to the Chern
number with
\begin{align} \label{eq:dC-gamma}
   \delta C^s_{\gamma} = (s/2)\;\mathrm{sign}(\Delta).
\end{align}

\subsubsection{Three-band crossings at $\kappa$ and $\kappa'$: $1.5$ Dirac cones} 

\begin{figure}[h]
\centering
\includegraphics[width=0.88\linewidth]{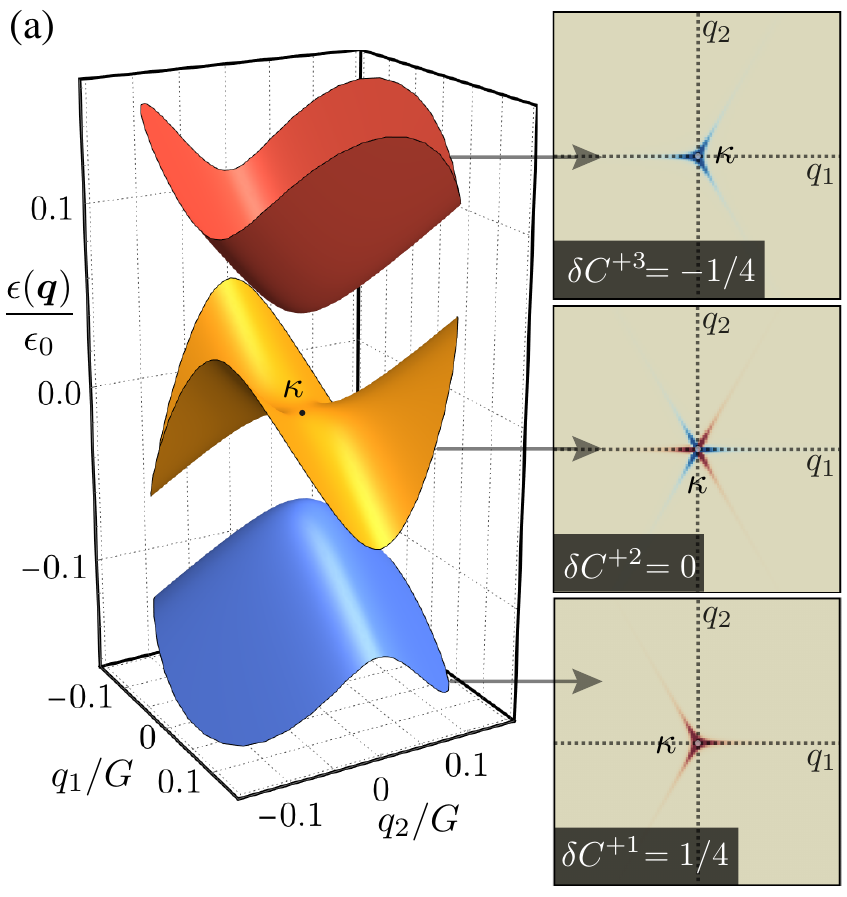}
\includegraphics[width=0.88\linewidth]{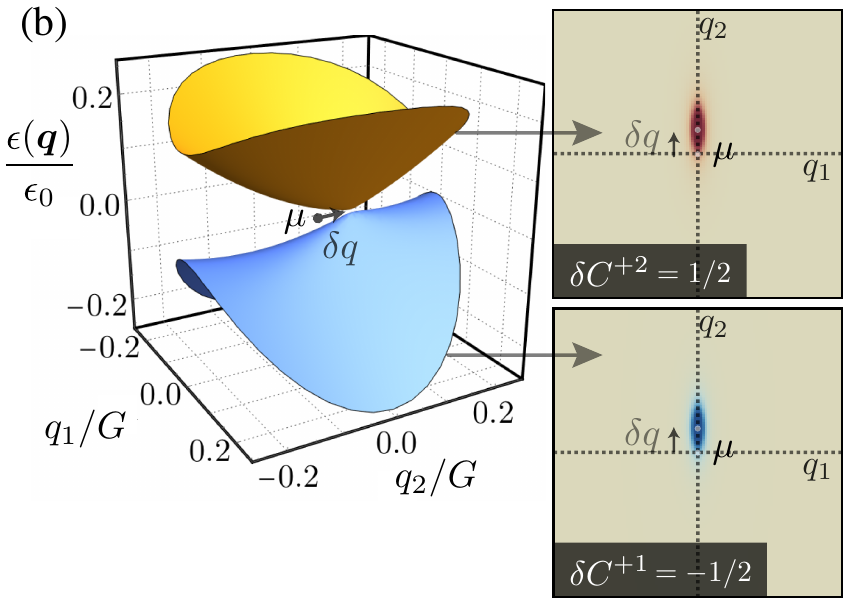}
\caption{\label{fig:BerryCurvatureLocal}
    Band structures and associated Berry curvatures calculated in a
    small-momentum expansion around the high-symmetry points. (a)~The 1.5
    Dirac cone, deriving from three cutting cones at the $\kappa$  or
    $\kappa'$ point (see Fig.\ \ref{fig:mBZ_hybridization2}), contribute with
    $\delta C_\zeta^{sn} = (\xi^{sn}/4)\, \mathrm{sign}( {\Delta_\zeta''}^s)$
    to the Berry curvature of the miniband; see also
    Ref.~[\onlinecite{Song2015}] for equivalent results found numerically.
    (b)~The two bands (deriving from two cutting cones; see Fig.\
    \ref{fig:mBZ_hybridization2}) approaching one another near the $\mu$  point
    contribute with $\delta C_\mu^{sn} = (\iota^{sn}/2)\,
    \mathrm{sign}(\tilde{\Delta})$ to the Berry curvature of the Bloch band.
}
\end{figure}

In order to determine the Berry curvatures near the $\kappa$  and
$\kappa'$ points, we study the three-band Hamiltonian 
\eqref{eq:BlochHamiltonian_zeta} for a purely TIAS situation. A
small-momentum expansion (with polar coordinates $\vec{q}=[q
\cos(\varphi),q\sin(\varphi)]$) around $\kappa$ or $\kappa'$ provides
the Hamiltonian
\begin{widetext}
\begin{align}
\label{eq:Hamiltonian_simple3band}
   H^s_\zeta(\vec{q}) \approx  \begin{pmatrix}
   \hbar v \, q \cos\left(\varphi\right) & \frac{i}{\sqrt{3}}\,\Delta'' 
   & -\frac{i}{\sqrt{3}}\,\Delta'' \\
   -\frac{i}{\sqrt{3}}\,\Delta'' & \hbar v \,q \cos\left(\varphi
   +\frac{2\pi}{3}\right) & \frac{i}{\sqrt{3}}\,\Delta'' \\ \frac{i}{\sqrt{3}}\,\Delta'' 
   & -\frac{i}{\sqrt{3}}\,\Delta'' 
   &  \hbar v \,q \cos\left(\varphi + \frac{4\pi}{3}\right)
   \end{pmatrix} ,
\end{align}
\end{widetext}
where $ \Delta'' = {\Delta^s_\zeta}''$ is the magnitude of the TIAS band
splitting and we choose the zero of energy at the high-symmetry point
$\epsilon^s(\vec\kappa)$, see Eq.\ \eqref{eq:Energies33_main0}. We restrict the
sums in \eqref{eq:BerryCurvature2D} to the three minibands constituting the 1.5
Dirac cone and make use of the eigenenergies and eigenfunctions of the
Hamiltonian \eqref{eq:Hamiltonian_simple3band}; see Appendix
\ref{sec:Eigensolutions}. With proper arrangement of terms and using
symmetries, one arrives at the Berry curvatures
\begin{align}
\label{eq:15DiracBerryCurvature}
   \Omega^n(\vec{q}) & \!=\! \frac{\hbar^2 \, v^2 \, (\frac{2}{\sqrt{3}}\Delta'')^{3}}
   {[(\frac{2}{\sqrt{3}}\Delta'')^2\!+\!(\hbar v)^2 \, q^2]^{5/2}} \; 
   \Phi\left[\phi(\vec{q})+{2\pi \iota^{n}}/{3}\right],
\end{align}
for the bands $n=1,2,3$ with $\phi(\vec{q})$ and $\iota^n$ defined above; see
Eq.\ \eqref{eq:Energies33_mainp} (we suppress the indices $\zeta$ and $s$).
The factor
\begin{align}
   \Phi(\phi) &= \frac{4}{\sqrt{3}} \frac{\cos(\phi)}{[1+\cos(2\phi)]^3}
\end{align}
describes a threefold $\varphi$-dependent modulation of the Berry curvature.
Furthermore, given the simpler form of Eq.\ \eqref{eq:Hamiltonian_simple3band},
the angle $\phi(\vec{q})$ reduces to 
\begin{align}
   \phi(q,\varphi) = \frac 13 \arccos\left[\frac{(\hbar v \, q)^3 
   \cos(3\varphi)}{[(\frac{2}{\sqrt{3}}\Delta'')^2+\hbar^2 v^2 \, q^2]^{3/2}}\right].
\end{align}
The integration of $\Phi[\phi(q,\varphi) + 2\pi \iota^n /3]$ over the angle
$\varphi$ generates the dependence of the Berry curvature on $n$,
\begin{align}\label{eq:int_varphi}
    \int_0^{2\pi}\dd{\varphi} \Phi\left(\phi_j(q,\varphi) \right) =
    \xi^n
    \bigl[\bigl(2\Delta^{\prime\prime}\bigr)^2/3+ (\hbar v)^2 \, q^2\bigr],
\end{align}
with $\xi^n = \mathrm{sign}\{\cos[(1+2\iota^{n})\pi/3]\} \in \{-1,0,1\}$.
As a result, the Berry curvature \eqref{eq:15DiracBerryCurvature} assumes the
form of a broadened (by $\Delta''/\hbar v$) and warped 2D $\delta$-function of
weight 1/4 (with the index $s$ reinstalled),
\begin{align}
   \delta C_\zeta^{sn} =  (\xi^{sn}/4)\; 
   \mathrm{sign}({\Delta^s_\zeta}'').
\end{align}
In Fig.~\ref{fig:BerryCurvatureLocal}, we show the dispersions of the three
bands $n = 1, 2, 3$ at the $\kappa$ point together with a Berry curvature map
exhibiting a $C_3$ symmetric angular dependence.  Such a Berry curvature map
with total weights $\pm 1/4$ and zero for the top/bottom and middle bands has
been found numerically~\cite{Song2015}. Our analytical result provides
additional insight on the origin of these weights: the distribution of the 1.5
Dirac cone Berry curvature can be understood as splitting the zero curvature at
degeneracy into two parts $1/2$ and $-1/2$ attributed to the top ($n = 2,3$)
and bottom ($n = 1, 2$) pairs of bands.  The weight $\pm 1/2$ is again equally
split between the pair constituents, such that the middle band ends up with a
zero integrated curvature $\delta C_\zeta^{s2} = 1/4 - 1/4 = 0$, while the top
and bottom bands remain with a weight $\delta C_\zeta^{s1} = \pm 1/4$ and
$\delta C_\zeta^{s3} = \mp 1/4$.

Above, we have discussed the novel situation of emergent Berry curvatures when
splitting the 1.5 Dirac cones at the $\kappa$  and $\kappa'$ points. A
different situation arises when the three bands crossing at these points is
first split with a finite TIS parameter, see Fig.\
\ref{fig:TriangleRotations}(b), leaving a singlet and a doublet at the $\kappa$
and $\kappa'$ points.  The doublet then is associated with a conventional
(warped) Dirac cone where a finite TIAS parameter frees a conventional Berry
curvature of integrated weight $\pm 1/2$.

\subsubsection{Two-band crossings near $\mu$: anisotropic Dirac cone}

The second nontrivial contribution to the Berry curvature arises from the
two-band splitting near the $\mu$  point; see Fig.\
\ref{fig:BerryCurvatureLocal}. While in  Fig.\
\ref{fig:PhaseSpace_Asymm}~\textbf{B} the two bands $n = 1,2$ (in ascending
order of excitation energy) cross in $\mu$, in a more general situation, these
bands split and free a Berry curvature that resides a finite distance $\delta
q$ away from the $\mu$  point; see Figs.\
\ref{fig:PhaseSpace_Asymm}~\textbf{A,~C,~D,~F} as well as Fig.\
\ref{fig:BerryCurvatureLocal}. In order to derive the Berry curvature for this
situation, we again employ a small-momentum expansion with $\vec{q}=(q_1,q_2)$
measured with respect to the point $\vec{\mu} = \vec{G}_0/2=(G/2,0)$ in
reciprocal space. Including both TIS and TIAS parameters, we find the
characteristic two-band anisotropic Dirac Hamiltonian
\begin{align}
\label{eq:Hamiltonian_Simple2Band}
   H^s_\mu(\vec{q}) = \begin{pmatrix}
   \hbar v\; (q_1 + q_2^2/G) & \frac12\bar\Delta + \hbar \bar{v}\, q_2\,e^{i\,\theta} \\
   \frac12{\bar\Delta}\strut^* + \hbar {\bar{v}}\, q_2\, e^{-i\,\theta} 
   & \hbar v\; (-q_1 + q_2^2/G)
\end{pmatrix},
\end{align}
where $G$ is again the reciprocal lattice constant and we chose the zero of
energy at $\epsilon^s(\vec{\mu})$. We see that the substrate potential defines
the gap $\bar{\Delta}$ and the renormalized velocity $\bar{v}$ along $q_2$ via
\cite{Wallbank2013}
\begin{align}
   \bar{\Delta}\!=\! 2s (\aS\!-\!i\,\aA)\! -\! 2(\mS\!-\!i\,\mA), 
   && \hbar \bar{v} G \,e^{i\,\theta} \!=\!  2(\uS\!+\!i\,\uA)\nonumber.
\end{align}
A straight-forward evaluation of Eq.~\eqref{eq:BerryCurvature2D} using the
eigenenergies and eigenfunctions of \eqref{eq:Hamiltonian_Simple2Band}, see
Appendix \ref{sec:Eigensolutions}, results in the Berry curvatures [with
$\iota^{sn} = s(2n-3) \in \{-1,1\}$]
\begin{align} \label{eq:Berry_Warped2band}
    \Omega^{sn}(\vec{q})   & =\frac{\iota^{sn}}2 \frac{\hbar v\, \hbar\bar{v} \,
   \tilde{\Delta}}{\bigl[{\tilde{\Delta}}\strut^2 \!+\! (\hbar v)^2\, q_1^2
   +(\hbar \bar{v})^2\,(q_2-\delta q)^2\bigr]^{3/2}},
\end{align}
where $\tilde{\Delta} = \tilde{\Delta}^s = (s \aS \!-\!  \mS)\,
\sin\theta-(s\aA\!-\!\mA)\, \cos\theta$ is the minimal gap displaced from the
$\mu$  point by $\hbar \bar{v}\, \delta q = (s\aA-\mA))\, \sin\theta - (s \aS -
\mS)\, \cos\theta$ along the $q_2$-direction, i.e., towards the $\kappa$  or
$\kappa'$ point. The Berry curvature Eq.\ \eqref{eq:Berry_Warped2band} assumes
the form of an anisotropically broadened 2D $\delta$-function with weight
$n/2$, see Fig.\ \ref{fig:BerryCurvatureLocal}; in the limit $\tilde\Delta \to
0$, it approaches the 2D $\delta$-function $\Omega^{sn} (\vec{q}) \to
(\iota^{sn}/2)\, \mathrm{sign}(\tilde\Delta)\, \delta(q_1) \delta(q_2-\delta
q)$. The integration of \eqref{eq:Berry_Warped2band} thus contributes a term
\begin{align}
   \delta C_\mu^{sn} =  (\iota^{sn}/2)\; \mathrm{sign}[\tilde\Delta^s].
\end{align}

As $\delta q$ becomes large, the curvatures at $\kappa$ and $q_2 -\delta q$
start overlapping and our approximations break down. Nevertheless, the
contributions still add, until a gap closure intervenes when $\delta q =
G/2\sqrt{3}$ (i.e., the minimal gap from $\mu$ passes through the
$\kappa$ point) and the Berry curvatures originating from the $\mu$  point get
redistributed between the second and third band.

The Chern number of the individual minibands is obtained by adding the
contributions from the $\gamma$ , $\kappa$ , $\kappa'$ , and $\mu$  points.
Focusing on the lowest electron  and hole bands (with $n = 1$), we first
define the contribution $\delta C^s = \delta C^{s}_\mu + \delta C_+^{s} +
\delta C_-^{s}$ arising from the vicinity of the $\kappa$ $\mu$  $\kappa'$
points, with $\delta C^{s}_\mu = \pm 3/2$ and $\delta C_\pm^{s} = \pm 1/4$ and
the factor 3 arising from the three $\mu$  points in the first BZ. These
contributions then add up to values $\delta C^s \in \{\pm 1/2,\pm 3/2 \}$ and
are shown in Fig.\ \ref{fig:PhaseSpace_Asymm} for the various maps \textbf{A}
-- \textbf{G}.  Adding the contribution $\delta C_\gamma^s = \pm 1/2$ from the
$\gamma$ point, we can reach values $C^s \in \{0, \pm 1, \pm 2\}$ for the
Chern numbers of the lowest excitation bands.

In the above discussion, we have focused on a single Dirac cone deriving,
e.g., from an original $K$-point; adding a time-reversed cone at $K'$ then
adds the same bands but with opposite Berry curvatures such that the total
Berry curvatures add up to zero (note that for a graphene derived system it is
always the inversion symmetry that breaks the TI symmetry). In this situation,
nontrivial topology manifests itself in valley
physics~\cite{Xiao2007,Rycerz2006}; only upon breaking of time-reversal
symmetry can the Berry curvatures of separate cones be decoupled and overall
topological effects be realized.

\section{Topological valley insulator from filled minibands}\label{sec:ins}

\begin{figure}[h]
    \includegraphics[width=8.6cm]{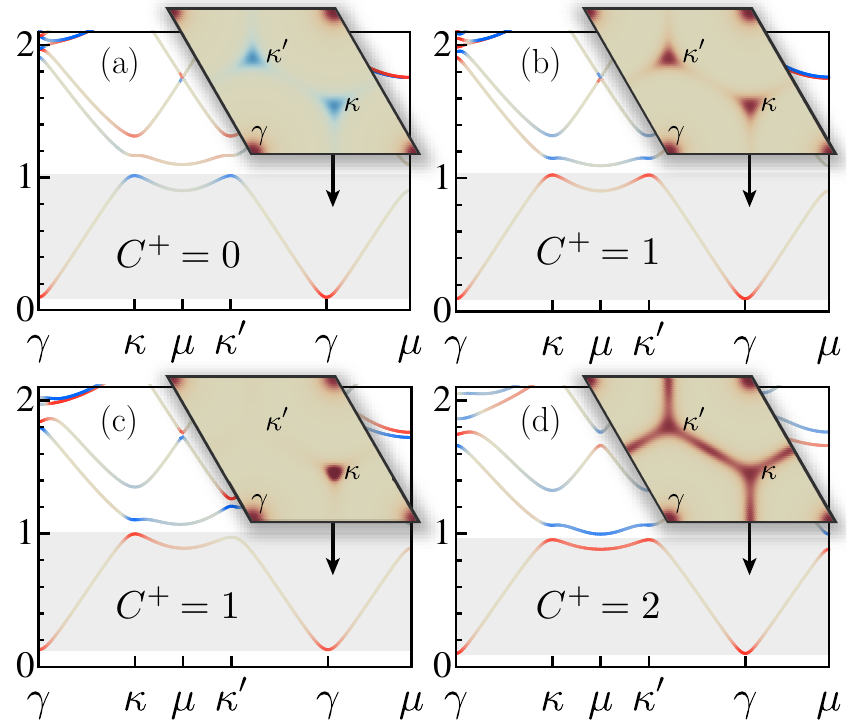} 
    \caption{\label{fig:special_configs}
	Band structures and Berry curvature maps realizing different
    topological insulator phases with valley Chern numbers $C^+=0,1,2$ for the
    lowest electronic band (light grey shading).  The local Berry curvatures
    are indicated in red and blue (see online colors) for positive and
    negative contributions, respectively. The Berry curvature is calculated
    using numerical diagonalization involving $62$ bands and the Chern number
    is obtained with the help of Fukui's method \cite{Fukui2007}.  The
    parameters describing the substrate potential for the cases {
    (a)--(d)} can be found in Table~\ref{tab:parameters}.  Cases (a)
    and (b) have been discussed in the context of graphene on
    hexagonal boron-nitride involving either smooth incommensurate Moir\'e
    structures (with $C^+ =0$) or commensurate grain boundaries ($C^+ = 1$)
    \cite{Song2015}. Case (c) highlights a situation where the
    Berry curvature has been pushed from a symmetric distribution to the
    $\kappa'$ point. Case (d) shows a Berry curvature network
    connecting all $\kappa$  and $\kappa'$ points and accumulating a 
    total valley Chern number $C^+ = 2$. 
    }
\end{figure}

In order to realize topological (valley) physics \cite{Xiao2007,Rycerz2006} with
surface-induced minibands, we have to tune the $3+3+1$ TIS, TIAS, and mass
parameters in \eqref{eq:ModelHamiltonian} such as to generate isolated bands
with finite Chern numbers and place the Fermi level in the minigap. While
numerous parameter settings provide access to such conditions, in Table
~\ref{tab:parameters} and Fig.~\ref{fig:special_configs} we present a few
illustrative examples of the kind of interesting physics that can be brought
forward.
\begin{table}
    \centering
    \begin{ruledtabular}
        \begin{tabular}{r rrr|rrrc|cc}
                            & \multicolumn{3}{c|}{\scriptsize (T+I)-symmetric}
                            & \multicolumn{3}{c}{\scriptsize
(T+I)-antisymmetric} &                  &                                      
\\
                            & $\uS$                                            
& $\mS$            & $\aS$  & $\uA$   & $\mA$   & $\aA$  & [$\epsilon_0$]
                            & $C^+$                                            
& $C^-$                                                                        

\\ \hline
            (a)   & $0$                                              
& $0$              & $0$    & $0$     & $-0.10$ & $0$    &
                            & 0                                                
& \phantom{$-$}$0$                                                             

\\
            (b)  & $0$                                              
& $0$              & $0$    & $0$     & $0.10$  & $0$    &                & 1 &
$-1$
\\
            (c) & $0$                                              
& $0$          & $0.1$ & $-0.05$ & $0$     & $0$ &                & 1 &
$-1$
\\
            (d)  & $0.20$                                           
& $0$              & $0$    & $0$  & $-0.12$ & $0$    &                & 2 &
$0$
        \end{tabular}    \end{ruledtabular}

    \caption{\label{tab:parameters}
        TIS and TIAS parameters for several topological band insulators
        involving the lowest electron  and hole-type minibands and resulting
        Chern numbers $C^+$ and $C^-$. The $\gamma$ point is gapped with a
        mass ${\Delta=0.1\, \epsilon_0 > 0}$. The associated Berry curvatures
        are shown in Fig.~\ref{fig:special_configs}.
    }
\end{table}

Cases (a) and (b) have been chosen with reference to the
work of Song {\it et al.}~\cite{Song2015} describing the emergence of
(topological) minibands in the G-hBN system. It turns out that a smooth and
incommensurate Moir\'e pattern with antisymmetric Dirac mass modulation $\mA$
generates a topologically trivial miniband: the homogeneous mass parameter
$\Delta$ derived from $\mA$ perturbatively in third order assumes a sign
different from $\mA$, see Eq.\ \eqref{eq:D3}, such that the Berry curvatures
at $\gamma$ cancel against those at the $\kappa$  and $\kappa'$ points; in our
phenomenological description this corresponds to the case (a) with
$\mA = - \Delta = - 0.1\, \epsilon_0$. On the contrary, modulating the
graphene with a commensurate grain boundary network~\cite{Song2015}
produces equal-sign masses that corresponds to our case (b).

In case (c), we have chosen parameters such as to move the Berry curvature
between $\kappa$  and $\kappa'$ points. We start from a situation with very
asymmetric band arrangement between the $\kappa$  and $\kappa'$ points as
driven by a large parameter $\aS$ (or alternatively $\mS$); see the
configuration {\bfseries G} (or {\bfseries E}) in Fig.\
\ref{fig:PhaseSpace_Symm}. The lowest two bands then have a single band
touching at $\kappa'$ and vanishing Berry curvature due to TI symmetry. The
gap opening at $\kappa'$ controlled by the doublet splitting
$\Delta_-^{+\,\prime\prime}\sim\uA$ frees a large Berry curvature at the
$\kappa'$ point. Note that with this procedure, we gap a conventional Dirac
cone (providing a contribution $\delta C_+^+ = 1/2$ from the $\kappa$ point)
rather than the 1.5 Dirac cone discussed above and apply to case (b)
(providing two contributions $\delta C_+^+ = 1/4$ and $\delta C_-^+ = 1/4$
from the $\kappa$  and $\kappa'$ points).

Finally, in case (d), we start from a situation with symmetric band
configurations at $\kappa$ and $\kappa'$ and a band crossing at $\mu$ as driven
by the large TIS parameter $\uS$; see the Fig.\ \ref{fig:PhaseSpace_Symm}~B.
Choosing a finite TIAS parameter $\mA$ frees the large Berry curvature near
the $\mu$  point.  The flatness of the bands along the $\kappa$ -$\mu$
-$\kappa'$ line spreads the Berry curvature to connect all $\kappa$ and
$\kappa'$ points and thereby generates the curvature network illustrated in
Fig.\ \ref{fig:special_configs} (d). Note that, with this choice of parameters,
the Berry curvature near the $\kappa$  and $\kappa'$ points derives from the
$\mu$  point and accumulates a total valley Chern number $C^+ = 2$. An
alternative realization of such a Berry curvature network with $C^+ = 2$ has
been found in Ref.\ [\onlinecite{SanJose2014a}] for spontaneously strained
graphene on hBN, however, without a protecting minigap. Within our formulation,
such a strain-induced situation is captured by an additional finite $\aA$
parameter, e.g., for parameters $\Delta=0.1\,\epsilon_0$,
$\uS=0.2\,\epsilon_0$, $\aA=0.12\,\epsilon_0$, we obtain a similar Berry
curvature network as in Fig.~\ref{fig:special_configs}(d) for an isolated band
with a Chern number $C^+ = -2$.

\section{Summary and Conclusion}\label{sec:concl}

In the present paper, we have subjected a Dirac-like particle to a periodic
substrate potential and have calculated the ensuing band structure as well as
its topological properties. Within our phenomenological approach, the model
Hamiltonian involves the Fermi velocity $v$ of the Dirac-like particle,
possibly a finite (TI-antisymmetric) mass $\Delta$ opening a gap at the
$\gamma$ point, and $3+3$ TI-symmetric (TIS) and TI-antisymmetric (TIAS)
parameters opening up gaps at the $\kappa$, $\kappa'$, and $\mu$  points. While
TIS parameters leave a cone at the $\kappa$ and $\kappa'$ points, these band
touchings are lifted by the TIAS parameters and a finite Berry curvature
emerges at the $\gamma$ (due to a finite $\Delta$) as well as at the $\kappa$,
$\kappa'$, and $\mu$  points (due to the TI-antisymmetric potential part). Such
a system opens the possibility for deliberate miniband engineering and tuning
of the Dirac material between different (valley) topological phases. 

The phenomenological Dirac-like model described in this paper involves a
single Dirac cone---in reality, such a model originates from a microscopic
bandstructure where the microscopic lattice generates the effective low-energy
Dirac-like dispersion, while the periodic substrate potential defines a
secondary or miniband structure. Microscopic time-reversal symmetry then
generates a partner valley that compensates the Berry curvatures of the
original Dirac cone. The topological properties of the Dirac material then
are reduced to valley-specific features that have to be brought to manifest
through special measures~\cite{Schaibley2016}, e.g., via exploiting the valley
Hall effect in a nonlocal conductance measurement as proposed in Refs.\
[\onlinecite{Xiao2007, Song2015}] and measured in Refs.\
[\onlinecite{Gorbachev2014}]. While this type of (bulk)
measurement probes the valley Berry curvature, other techniques, such as
scanning tunneling spectroscopy, attempt to image topological edge states; this
latter technique has been successfully applied to states associated with the
main gap in graphene bilayers~\cite{Yin2016,Huang2018}. In photonic systems~\cite{Ozawa2018} designing the edge and boundary termination is easier such that (valley) edge states can be imaged directly~\cite{Ma2016,Dong2017,Kang2018,Noh2018}.

The realization of valley-topological physics through substrate-assisted
miniband engineering involves proper tuning in a high-dimensional parameter
space. Not only does one require a proper set of TIAS parameters bringing
forward Berry curvatures with finite Chern numbers, in addition, the
topological minibands have to be isolated from the other bands through proper
gaps. Our analysis of TIS and TIAS parameter-spaces summarized in Figs.\
\ref{fig:PhaseSpace_Symm} and \ref{fig:PhaseSpace_Asymm} provides a systematic
overview of possible arrangements of miniband structures for the lowest bands
in the vicinity of the $\kappa$, $\kappa'$, and $\mu$  points. Our specific
examples in Fig.\ \ref{fig:special_configs} demonstrate that isolated mini-
bands with nontrivial Chern numbers can be achieved in principle. The next
important step in a program aiming at substrate-assisted topological mini- band
engineering then has to establish the connection between the phenomenological
and microscopic parameters describing real band electrons subject to a
substrate potential. Inspiration for the solution of this task can be gained
from several principles: the TIAS mass parameter $\Delta$ derives from
sublattice asymmetry \cite{Zhou2007} breaking inversion symmetry, the potential
parameters $u$ can be engineered with an electrostatic top-gate
pattern~\cite{Park2008a,Ye2011}, finite mass parameters $m$ derive from a
modulated sublattice asymmetry \cite{Wallbank2013,Moon2014}, and gauge field
parameters $a$ are induced by bond modulations, e.g., through strain~\cite{CastroNeto2009,SanJose2014a,Jung2015}. More detailed analysis then
involves realistic bandstructure calculations that pose a challenging problem
given the large supercell of Moir\'e or grainboundary structures in real
systems.  Alternatively, optically engineered atomic
crystals~\cite{Uehlinger2013,Jotzu2014} and optical waveguide
arrays~\cite{Ozawa2018,Noh2018} may provide another arena for the implementation
of topological minibands.

\section*{Acknowledgments}

We thank M.\ S.\ Ferguson, J.\ Lado, M.\ H.\ Fischer, and I.\ Petrides for illuminating
discussions and acknowledge financial support from the Swiss National Science
Foundation, Division 2 and through the National Centre of Competence in
Research ``QSIT - Quantum Science and Technology''.


\appendix 

\section{Local $U(1)$ symmetry and transverse gauge}
\label{sec:tr_gauge}

For the reader's convenience, we briefly elaborate on the role of gauge freedom
and gauge fixing in the model of a Dirac-like particle elastically scattered on
a static potential \cite{Wallbank2013}. For this purpose, let us consider the
Dirac particle minimally coupled to a vector potential
$\vec{A}(\vec{x})=(A_1(\vec{x}),\,A_2(\vec{x}))$, i.e.,
\begin{align}
H = {v}\,\left[\hbar\vec{k}+\vec{A}(\vec{x})\right]\cdot\vec{\sigma} +
\Delta\,\sigma_3 \,,
\end{align}
where ${v}$, $\vec{k}$, $\vec{\sigma}$ and $\Delta$ are defined in the main text,
see Eq.~\eqref{eq:BlochHamiltonian}. The vector potential can always be
decomposed into longitudinal and transverse parts, i.e.,
$\vec{A}(\vec{x}) = \vec{A}_\parallel(\vec{x}) + \vec{A}_\perp(\vec{x})$ with
$\vec{\nabla}\times\vec{A}_\parallel(\vec{x})=0$ and
$\vec{\nabla}\cdot\vec{A}_\perp(\vec{x})=0$. We can furthermore express these
components through scalar functions, i.e.,
\begin{align}
 \vec{A}_\parallel(\vec{x}) = \nabla a_\parallel(\vec{x}),
 && \vec{A}_\perp(\vec{x}) = \hat{z}\times\nabla a_\perp(\vec{x}) \; .
\end{align}
The vector potential $\vec{A}(\vec{x})$ can have different physical origins,
such as the presence of electromagnetic fields. However, in G-hBN it arises due
to the proximity of the substrate layer even in absence of external fields. The
(pseudo) magnetic field associated with the vector potential is given by
\begin{align}
\vec{B}_{\text{pseudo}}=\nabla\times\vec{A}(\vec{x})
    = \nabla^2a_\perp(\vec{x}) \; \vec{\hat{z}} \; .
\end{align}

We are allowed to make the unitary transformation $U(\vec{x}) =
e^{-i\,a_\parallel(\vec{x})}$ without changing the physics of our system. This
is our local $\mathrm{U}(1)$ gauge freedom. The identity
\begin{align}
U^\dagger(\vec{x})\,[-i \vec{\nabla}]\cdot\vec{\sigma}\,U(\vec{x}) = [-i
\vec{\nabla}-\nabla a_\parallel(\vec{x})]\cdot\vec{\sigma}
\end{align}
then readily implies that the transformation $U(\vec{x})$ removes the
longitudinal component $\vec{A}_\parallel(\vec{x})$ of the vector potential
from the Hamiltonian,
\begin{align}
U^\dagger(\vec{x}) \, H \, U(\vec{x}) &=
{v}\,\left[\hbar\vec{k}+\vec{A}_\perp(\vec{x})\right]\cdot\vec{\sigma} +
\Delta\,\sigma_3  \\ &= \left[{v}\,\hbar\vec{k}\cdot\vec{\sigma} +
\Delta\,\sigma_3\right] + \hbar{v} \; \hat{z}\times \nabla a_\perp(\vec{x}) \;
. \nonumber
\end{align}

We conclude that whenever a term of the form $\vec{A}(\vec{x}) \cdot
\vec{\sigma}$ is present in the Hamiltonian of spinless non-interacting
Dirac-like particles, we can remove its longitudinal part through a suitable
gauge transformation.

\section{Topological minibands}
\label{sec:Eigensolutions}

\subsection{2D Abelian Berry curvature} 

For a given Hamiltonian $H(\vec{k})$ parametrized by the crystal momentum
$\vec{k}=(k_1,k_2)$ with eigenfunctions $\ket{n(\vec{k})}$ and eigenenergies
$\epsilon^n(\vec{k})$, we can define the 2D Abelian Berry curvature as
\begin{align}
\Omega^{n}(\vec{k}) &= [\vec{\nabla}\times \vec{A}^{(n)}(\vec{k})]_3 \nonumber \\
&=  \left[\vec{\nabla}\times \braket{n(\vec{k}) | \vec{\nabla} | n(\vec{k})}\right]_3 \nonumber \\
&= \partial_1 \braket{n(\vec{k}) | \partial_2 | n(\vec{k})} -  \partial_2 \braket{n(\vec{k}) | \partial_1 | n(\vec{k})} .
\end{align}
However, a major drawback of this expression is that it involves taking
derivatives of the wavefunctions. In practice, this way of calculating the
Berry curvature quickly becomes intractable. In some cases, this difficulty
can be overcome by trading the derivatives for sums over all bands containing
matrix elements with derivatives of the Hamiltonian and an energy denominator,
\cite{Xiao2010}
\begin{align}
   \Omega^{n}(\vec{k}) 
    &= i \sum_{n'\neq n}\frac{\braket{n|\partial_{k_1} H|n'}
    \braket{n'|\partial_{k_2} H|n}-\left(1\leftrightarrow 2\right)}
    {[\epsilon^{n}-\epsilon^{n'}]^2},
\end{align}
where we have suppressed the $\vec{k}$ dependence on the right-hand side in
our notation and we have made use of the identity $\braket{n|\partial_{k_j}
H(\vec{k})|n'} = \braket{\partial_{k_j}\,n|n'}(\epsilon_n-\epsilon_{n'})$ for
$n\neq n'$.

\subsection{Three-band crossing: 1.5 Dirac cone}

The three energy bands ($n=1,2,3$) of the effective Hamiltonian
\eqref{eq:Hamiltonian_simple3band} describing a fully hybridized three-band 
crossing can be obtained by solving the cubic characteristic equation. In this
case, we find
\begin{align}
   \epsilon^n(q,\varphi) &= \sqrt{(2\Delta'')^2/3
   + (\hbar v\,q)^2}\,\cos(\phi^n(q,\varphi)) , \nonumber\\
   \phi^n(q,\varphi) &= \frac 13 \arccos\left[\frac{[\hbar v\,q)^3 
   \cos(3\varphi)}{((2\Delta'')^2/3\!+\!(\hbar v\,q)^2]^{3/2}}\right] \!+\! 2\pi \iota^n/3 ,
\end{align}
where $\iota^n = s(n-1/2) + 1/2$ (as in the main text).
The associated unnormalized eigenvectors can be found to be
\begin{align}
\ket{n(q,\varphi)} = \begin{pmatrix}
1 + \chi^n(q,\varphi+2\pi/3) \\
1 + \chi^n(q,\varphi) \, \chi^\iota(q,\varphi+2\pi/3) \\
1 - \chi^n(q,\varphi)
\end{pmatrix},
\end{align}
where $\chi^n(q,\varphi)=i \, \left[ \hbar v\,q\,\cos(\varphi) -
\epsilon^n(q,\varphi)\right]/(\Delta''/\sqrt{3})$. The Berry curvature
\eqref{eq:15DiracBerryCurvature} for each band can then be obtained by
evaluating Eq.~\eqref{eq:BerryCurvature2D}, which requires some care in how to
arrange the various terms and the exploitation of symmetries.

\subsection{Two-band crossing: Anisotropic Dirac cone}

A simple two-spinor rotation
\begin{align}
   U = e^{i (\pi/2+\theta)\,\sigma_3/2}\cdot e^{i (\pi/2)\,\sigma_2/2} 
\end{align}
and substraction of $\hbar v \, q_2^2/G$ (irrelevant for the discussion of band
topology) brings the effective Hamiltonian \eqref{eq:Hamiltonian_simple3band}
into a form where the anisotropic Dirac Hamiltonian becomes manifest, 
\small
\begin{align}
\label{eq:AnisotropicDiracHamiltonian}
   U^\dagger H(\vec{q}) U = 
\begin{bmatrix}
\tilde{\Delta} & \!\!\!\!\!\!\! \hbar v\,q_1 - i \, \hbar \bar{v}\,(q_2\!-\!\delta q) \\
\hbar v\,q_1 + i \, \hbar \bar{v}\,(q_2\!-\!\delta q) & -\tilde{\Delta}
\end{bmatrix},
\end{align}
\normalsize
where $\theta$, $\tilde{\Delta}$, $\bar{v}$ and $\delta q$ are defined in
the main text before and after Eq.~\eqref{eq:Berry_Warped2band}. The
corresponding energy bands ($n=1,2$) can be readily found to be
\begin{align}
   \epsilon^{n}(\vec{q}) = \iota^n \sqrt{\tilde{\Delta}^2 
   \!+\! (\hbar v \,q_1)^2\!+\!(\hbar \bar{v})^2(q_2-\delta q)^2} ,
\end{align}
where $\iota^{n} = s(2n-3)$ (as in the main text).
In the rotated basis of Eq.~\eqref{eq:AnisotropicDiracHamiltonian}, the
associated normalized eigenvectors are
\begin{align}
\ket{n(\vec{q})} = \frac{1}{\sqrt{2}} \begin{pmatrix}
  \phantom{\pm}\sqrt{1+\tilde{\Delta}/\epsilon^{\pm}(\vec{q})} \, e^{-i\,\varphi(\vec{q})/2} \\
  \iota^n \sqrt{1-\tilde{\Delta}/\epsilon^{\pm}(\vec{q})} \, e^{i\,\varphi(\vec{q})/2\phantom{-}} 
\end{pmatrix},
\end{align}
where $\varphi(\vec{q}) = \mathrm{Arg}(q_1 + i \, (\bar{v}/v) q_2)$, similar to Eqs.~\eqref{eq:disp}~and~\eqref{eq:DiracEigenstates}. The Berry curvature
\eqref{eq:Berry_Warped2band} for an anisotropic cone can then be obtained by
evaluating Eq.~\eqref{eq:BerryCurvature2D}.


\bibliography{references}

\begin{thebibliography}{64}%
\makeatletter
\providecommand \@ifxundefined [1]{%
 \@ifx{#1\undefined}
}%
\providecommand \@ifnum [1]{%
 \ifnum #1\expandafter \@firstoftwo
 \else \expandafter \@secondoftwo
 \fi
}%
\providecommand \@ifx [1]{%
 \ifx #1\expandafter \@firstoftwo
 \else \expandafter \@secondoftwo
 \fi
}%
\providecommand \natexlab [1]{#1}%
\providecommand \enquote  [1]{``#1''}%
\providecommand \bibnamefont  [1]{#1}%
\providecommand \bibfnamefont [1]{#1}%
\providecommand \citenamefont [1]{#1}%
\providecommand \href@noop [0]{\@secondoftwo}%
\providecommand \href [0]{\begingroup \@sanitize@url \@href}%
\providecommand \@href[1]{\@@startlink{#1}\@@href}%
\providecommand \@@href[1]{\endgroup#1\@@endlink}%
\providecommand \@sanitize@url [0]{\catcode `\\12\catcode `\$12\catcode
  `\&12\catcode `\#12\catcode `\^12\catcode `\_12\catcode `\%12\relax}%
\providecommand \@@startlink[1]{}%
\providecommand \@@endlink[0]{}%
\providecommand \url  [0]{\begingroup\@sanitize@url \@url }%
\providecommand \@url [1]{\endgroup\@href {#1}{\urlprefix }}%
\providecommand \urlprefix  [0]{URL }%
\providecommand \Eprint [0]{\href }%
\providecommand \doibase [0]{https://doi.org/}%
\providecommand \selectlanguage [0]{\@gobble}%
\providecommand \bibinfo  [0]{\@secondoftwo}%
\providecommand \bibfield  [0]{\@secondoftwo}%
\providecommand \translation [1]{[#1]}%
\providecommand \BibitemOpen [0]{}%
\providecommand \bibitemStop [0]{}%
\providecommand \bibitemNoStop [0]{.\EOS\space}%
\providecommand \EOS [0]{\spacefactor3000\relax}%
\providecommand \BibitemShut  [1]{\csname bibitem#1\endcsname}%
\let\auto@bib@innerbib\@empty
\bibitem [{\citenamefont {{Von Klitzing}}(1986)}]{Klitzing1986}%
  \BibitemOpen
  \bibfield  {author} {\bibinfo {author} {\bibfnamefont {K.}~\bibnamefont {{Von
  Klitzing}}},\ }\bibfield  {title} {\bibinfo {title} {{The quantized Hall
  effect}},\ }\href {https://doi.org/10.1103/RevModPhys.58.519} {\bibfield
  {journal} {\bibinfo  {journal} {Rev. Mod. Phys.}\ }\textbf {\bibinfo {volume}
  {58}},\ \bibinfo {pages} {519} (\bibinfo {year} {1986})}\BibitemShut
  {NoStop}%
\bibitem [{\citenamefont {Kane}\ and\ \citenamefont {Mele}(2005)}]{kane2005}%
  \BibitemOpen
  \bibfield  {author} {\bibinfo {author} {\bibfnamefont {C.~L.}\ \bibnamefont
  {Kane}}\ and\ \bibinfo {author} {\bibfnamefont {E.~J.}\ \bibnamefont
  {Mele}},\ }\bibfield  {title} {\bibinfo {title} {{$\mathrm{Z}_2$ topological
  order and the quantum spin Hall effect}},\ }\href
  {https://doi.org/10.1103/PhysRevLett.95.146802} {\bibfield  {journal}
  {\bibinfo  {journal} {Phys. Rev. Lett.}\ }\textbf {\bibinfo {volume} {95}},\
  \bibinfo {pages} {146802} (\bibinfo {year} {2005})}\BibitemShut {NoStop}%
\bibitem [{\citenamefont {Hasan}\ and\ \citenamefont {Kane}(2010)}]{Hasan2010}%
  \BibitemOpen
  \bibfield  {author} {\bibinfo {author} {\bibfnamefont {M.~Z.}\ \bibnamefont
  {Hasan}}\ and\ \bibinfo {author} {\bibfnamefont {C.~L.}\ \bibnamefont
  {Kane}},\ }\bibfield  {title} {\bibinfo {title} {{Colloquium : Topological
  insulators}},\ }\href {https://doi.org/10.1103/RevModPhys.82.3045} {\bibfield
   {journal} {\bibinfo  {journal} {Rev. Mod. Phys.}\ }\textbf {\bibinfo
  {volume} {82}},\ \bibinfo {pages} {3045} (\bibinfo {year}
  {2010})}\BibitemShut {NoStop}%
\bibitem [{\citenamefont {Qi}\ and\ \citenamefont {Zhang}(2011)}]{Qi2011}%
  \BibitemOpen
  \bibfield  {author} {\bibinfo {author} {\bibfnamefont {X.~L.}\ \bibnamefont
  {Qi}}\ and\ \bibinfo {author} {\bibfnamefont {S.~C.}\ \bibnamefont {Zhang}},\
  }\bibfield  {title} {\bibinfo {title} {{Topological insulators and
  superconductors}},\ }\bibfield  {journal} {\bibinfo  {journal} {Rev. Mod.
  Phys.}\ }\textbf {\bibinfo {volume} {83}},\ \href
  {https://doi.org/10.1103/RevModPhys.83.1057} {10.1103/RevModPhys.83.1057}
  (\bibinfo {year} {2011}),\ \Eprint {https://arxiv.org/abs/1008.2026}
  {arXiv:1008.2026} \BibitemShut {NoStop}%
\bibitem [{\citenamefont {Haldane}\ and\ \citenamefont
  {Raghu}(2008)}]{Haldane2008}%
  \BibitemOpen
  \bibfield  {author} {\bibinfo {author} {\bibfnamefont {F.~D.~M.}\
  \bibnamefont {Haldane}}\ and\ \bibinfo {author} {\bibfnamefont
  {S.}~\bibnamefont {Raghu}},\ }\bibfield  {title} {\bibinfo {title} {{Possible
  realization of directional optical waveguides in photonic crystals with
  broken time-reversal symmetry}},\ }\href
  {https://doi.org/10.1103/PhysRevLett.100.013904} {\bibfield  {journal}
  {\bibinfo  {journal} {Phys. Rev. Lett.}\ }\textbf {\bibinfo {volume} {100}},\
  \bibinfo {pages} {1} (\bibinfo {year} {2008})}\BibitemShut {NoStop}%
\bibitem [{\citenamefont {Khanikaev}\ \emph {et~al.}(2012)\citenamefont
  {Khanikaev}, \citenamefont {{Hossein Mousavi}}, \citenamefont {Tse},
  \citenamefont {Kargarian}, \citenamefont {MacDonald},\ and\ \citenamefont
  {Shvets}}]{Khanikaev2012}%
  \BibitemOpen
  \bibfield  {author} {\bibinfo {author} {\bibfnamefont {A.~B.}\ \bibnamefont
  {Khanikaev}}, \bibinfo {author} {\bibfnamefont {S.}~\bibnamefont {{Hossein
  Mousavi}}}, \bibinfo {author} {\bibfnamefont {W.-K.}\ \bibnamefont {Tse}},
  \bibinfo {author} {\bibfnamefont {M.}~\bibnamefont {Kargarian}}, \bibinfo
  {author} {\bibfnamefont {A.~H.}\ \bibnamefont {MacDonald}},\ and\ \bibinfo
  {author} {\bibfnamefont {G.}~\bibnamefont {Shvets}},\ }\bibfield  {title}
  {\bibinfo {title} {{Photonic topological insulators}},\ }\href
  {https://doi.org/10.1038/nmat3520} {\bibfield  {journal} {\bibinfo  {journal}
  {Nat. Mater.}\ }\textbf {\bibinfo {volume} {12}},\ \bibinfo {pages} {233}
  (\bibinfo {year} {2012})}\BibitemShut {NoStop}%
\bibitem [{\citenamefont {Ozawa}\ \emph {et~al.}(2018)\citenamefont {Ozawa},
  \citenamefont {Price}, \citenamefont {Amo}, \citenamefont {Goldman},
  \citenamefont {Hafezi}, \citenamefont {Lu}, \citenamefont {Rechtsman},
  \citenamefont {Schuster}, \citenamefont {Simon}, \citenamefont {Zilberberg},\
  and\ \citenamefont {Carusotto}}]{Ozawa2018}%
  \BibitemOpen
  \bibfield  {author} {\bibinfo {author} {\bibfnamefont {T.}~\bibnamefont
  {Ozawa}}, \bibinfo {author} {\bibfnamefont {H.~M.}\ \bibnamefont {Price}},
  \bibinfo {author} {\bibfnamefont {A.}~\bibnamefont {Amo}}, \bibinfo {author}
  {\bibfnamefont {N.}~\bibnamefont {Goldman}}, \bibinfo {author} {\bibfnamefont
  {M.}~\bibnamefont {Hafezi}}, \bibinfo {author} {\bibfnamefont
  {L.}~\bibnamefont {Lu}}, \bibinfo {author} {\bibfnamefont {M.}~\bibnamefont
  {Rechtsman}}, \bibinfo {author} {\bibfnamefont {D.}~\bibnamefont {Schuster}},
  \bibinfo {author} {\bibfnamefont {J.}~\bibnamefont {Simon}}, \bibinfo
  {author} {\bibfnamefont {O.}~\bibnamefont {Zilberberg}},\ and\ \bibinfo
  {author} {\bibfnamefont {I.}~\bibnamefont {Carusotto}},\ }\bibfield  {title}
  {\bibinfo {title} {{Topological Photonics}},\ }\href
  {http://arxiv.org/abs/1802.04173} {\  (\bibinfo {year} {2018})},\ \Eprint
  {https://arxiv.org/abs/1802.04173} {arXiv:1802.04173} \BibitemShut {NoStop}%
\bibitem [{\citenamefont {Prodan}\ and\ \citenamefont
  {Prodan}(2009)}]{Prodan2009}%
  \BibitemOpen
  \bibfield  {author} {\bibinfo {author} {\bibfnamefont {E.}~\bibnamefont
  {Prodan}}\ and\ \bibinfo {author} {\bibfnamefont {C.}~\bibnamefont
  {Prodan}},\ }\bibfield  {title} {\bibinfo {title} {{Topological phonon modes
  and their role in dynamic instability of microtubules}},\ }\href
  {https://doi.org/10.1103/PhysRevLett.103.248101} {\bibfield  {journal}
  {\bibinfo  {journal} {Phys. Rev. Lett.}\ }\textbf {\bibinfo {volume} {103}},\
  \bibinfo {pages} {1} (\bibinfo {year} {2009})},\ \Eprint
  {https://arxiv.org/abs/0909.3492} {arXiv:0909.3492} \BibitemShut {NoStop}%
\bibitem [{\citenamefont {S{\"{u}}sstrunk}\ and\ \citenamefont
  {Huber}(2016)}]{Susstrunk2016}%
  \BibitemOpen
  \bibfield  {author} {\bibinfo {author} {\bibfnamefont {R.}~\bibnamefont
  {S{\"{u}}sstrunk}}\ and\ \bibinfo {author} {\bibfnamefont {S.~D.}\
  \bibnamefont {Huber}},\ }\bibfield  {title} {\bibinfo {title}
  {{Classification of topological phonons in linear mechanical
  metamaterials}},\ }\href {https://doi.org/10.1073/pnas.1605462113} {\bibfield
   {journal} {\bibinfo  {journal} {Proc. Natl. Acad. Sci.}\ }\textbf {\bibinfo
  {volume} {113}},\ \bibinfo {pages} {E4767} (\bibinfo {year} {2016})},\
  \Eprint {https://arxiv.org/abs/1604.01033} {1604.01033} \BibitemShut
  {NoStop}%
\bibitem [{\citenamefont {Haldane}(1988)}]{haldane1988}%
  \BibitemOpen
  \bibfield  {author} {\bibinfo {author} {\bibfnamefont {F.~D.~M.}\
  \bibnamefont {Haldane}},\ }\bibfield  {title} {\bibinfo {title} {{Model for a
  quantum Hall effect without Landau levels: Condensed-matter realization of
  the``parity anomaly''}},\ }\href
  {https://doi.org/10.1103/PhysRevLett.61.2015} {\bibfield  {journal} {\bibinfo
   {journal} {Phys. Rev. Lett.}\ }\textbf {\bibinfo {volume} {61}},\ \bibinfo
  {pages} {2015} (\bibinfo {year} {1988})}\BibitemShut {NoStop}%
\bibitem [{\citenamefont {Oka}\ and\ \citenamefont {Aoki}(2009)}]{Oka2009}%
  \BibitemOpen
  \bibfield  {author} {\bibinfo {author} {\bibfnamefont {T.}~\bibnamefont
  {Oka}}\ and\ \bibinfo {author} {\bibfnamefont {H.}~\bibnamefont {Aoki}},\
  }\bibfield  {title} {\bibinfo {title} {{Photovoltaic Hall effect in
  graphene}},\ }\href {https://doi.org/10.1103/PhysRevB.79.081406} {\bibfield
  {journal} {\bibinfo  {journal} {Phys. Rev. B - Condens. Matter Mater. Phys.}\
  }\textbf {\bibinfo {volume} {79}},\ \bibinfo {pages} {1} (\bibinfo {year}
  {2009})},\ \Eprint {https://arxiv.org/abs/0807.4767} {arXiv:0807.4767}
  \BibitemShut {NoStop}%
\bibitem [{\citenamefont {Rechtsman}\ \emph {et~al.}(2013)\citenamefont
  {Rechtsman}, \citenamefont {Zeuner}, \citenamefont {Plotnik}, \citenamefont
  {Lumer}, \citenamefont {Podolsky}, \citenamefont {Dreisow}, \citenamefont
  {Nolte}, \citenamefont {Segev},\ and\ \citenamefont
  {Szameit}}]{Rechtsman2013}%
  \BibitemOpen
  \bibfield  {author} {\bibinfo {author} {\bibfnamefont {M.~C.}\ \bibnamefont
  {Rechtsman}}, \bibinfo {author} {\bibfnamefont {J.~M.}\ \bibnamefont
  {Zeuner}}, \bibinfo {author} {\bibfnamefont {Y.}~\bibnamefont {Plotnik}},
  \bibinfo {author} {\bibfnamefont {Y.}~\bibnamefont {Lumer}}, \bibinfo
  {author} {\bibfnamefont {D.}~\bibnamefont {Podolsky}}, \bibinfo {author}
  {\bibfnamefont {F.}~\bibnamefont {Dreisow}}, \bibinfo {author} {\bibfnamefont
  {S.}~\bibnamefont {Nolte}}, \bibinfo {author} {\bibfnamefont
  {M.}~\bibnamefont {Segev}},\ and\ \bibinfo {author} {\bibfnamefont
  {A.}~\bibnamefont {Szameit}},\ }\bibfield  {title} {\bibinfo {title}
  {{Photonic Floquet topological insulators}},\ }\href
  {https://doi.org/10.1038/nature12066} {\bibfield  {journal} {\bibinfo
  {journal} {Nature}\ }\textbf {\bibinfo {volume} {496}},\ \bibinfo {pages}
  {196} (\bibinfo {year} {2013})},\ \Eprint {https://arxiv.org/abs/1211.5623}
  {1211.5623} \BibitemShut {NoStop}%
\bibitem [{\citenamefont {Jotzu}\ \emph {et~al.}(2014)\citenamefont {Jotzu},
  \citenamefont {Messer}, \citenamefont {Desbuquois}, \citenamefont {Lebrat},
  \citenamefont {Uehlinger}, \citenamefont {Greif},\ and\ \citenamefont
  {Esslinger}}]{Jotzu2014}%
  \BibitemOpen
  \bibfield  {author} {\bibinfo {author} {\bibfnamefont {G.}~\bibnamefont
  {Jotzu}}, \bibinfo {author} {\bibfnamefont {M.}~\bibnamefont {Messer}},
  \bibinfo {author} {\bibfnamefont {R.}~\bibnamefont {Desbuquois}}, \bibinfo
  {author} {\bibfnamefont {M.}~\bibnamefont {Lebrat}}, \bibinfo {author}
  {\bibfnamefont {T.}~\bibnamefont {Uehlinger}}, \bibinfo {author}
  {\bibfnamefont {D.}~\bibnamefont {Greif}},\ and\ \bibinfo {author}
  {\bibfnamefont {T.}~\bibnamefont {Esslinger}},\ }\bibfield  {title} {\bibinfo
  {title} {{Experimental realization of the topological Haldane model with
  ultracold fermions}},\ }\href {http://dx.doi.org/10.1038/nature13915
  http://10.0.4.14/ nature13915
  http://www.nature.com/nature/journal/v515/n7526/
  abs/nature13915.html#supplementary-information} {\bibfield  {journal}
  {\bibinfo  {journal} {Nature}\ }\textbf {\bibinfo {volume} {515}},\ \bibinfo
  {pages} {237} (\bibinfo {year} {2014})}\BibitemShut {NoStop}%
\bibitem [{\citenamefont {Wu}\ and\ \citenamefont {Hu}(2015)}]{Wu2015}%
  \BibitemOpen
  \bibfield  {author} {\bibinfo {author} {\bibfnamefont {L.-H.}\ \bibnamefont
  {Wu}}\ and\ \bibinfo {author} {\bibfnamefont {X.}~\bibnamefont {Hu}},\
  }\bibfield  {title} {\bibinfo {title} {{Scheme for Achieving a Topological
  Photonic Crystal by Using Dielectric Material}},\ }\href
  {https://doi.org/10.1103/PhysRevLett.114.223901} {\bibfield  {journal}
  {\bibinfo  {journal} {Phys. Rev. Lett.}\ }\textbf {\bibinfo {volume} {114}},\
  \bibinfo {pages} {223901} (\bibinfo {year} {2015})}\BibitemShut {NoStop}%
\bibitem [{\citenamefont {Rycerz}\ \emph {et~al.}(2007)\citenamefont {Rycerz},
  \citenamefont {Tworzyd{\l}o},\ and\ \citenamefont {Beenakker}}]{Rycerz2006}%
  \BibitemOpen
  \bibfield  {author} {\bibinfo {author} {\bibfnamefont {A.}~\bibnamefont
  {Rycerz}}, \bibinfo {author} {\bibfnamefont {J.}~\bibnamefont
  {Tworzyd{\l}o}},\ and\ \bibinfo {author} {\bibfnamefont {C.~W.}\ \bibnamefont
  {Beenakker}},\ }\bibfield  {title} {\bibinfo {title} {{Valley filter and
  valley valve in graphene}},\ }\href {https://doi.org/10.1038/nphys547}
  {\bibfield  {journal} {\bibinfo  {journal} {Nat. Phys.}\ }\textbf {\bibinfo
  {volume} {3}},\ \bibinfo {pages} {172} (\bibinfo {year} {2007})},\ \Eprint
  {https://arxiv.org/abs/0608533} {0608533} \BibitemShut {NoStop}%
\bibitem [{\citenamefont {Xiao}\ \emph {et~al.}(2007)\citenamefont {Xiao},
  \citenamefont {Yao},\ and\ \citenamefont {Niu}}]{Xiao2007}%
  \BibitemOpen
  \bibfield  {author} {\bibinfo {author} {\bibfnamefont {D.}~\bibnamefont
  {Xiao}}, \bibinfo {author} {\bibfnamefont {W.}~\bibnamefont {Yao}},\ and\
  \bibinfo {author} {\bibfnamefont {Q.}~\bibnamefont {Niu}},\ }\bibfield
  {title} {\bibinfo {title} {{Valley-contrasting physics in graphene: Magnetic
  moment and topological transport}},\ }\href
  {https://doi.org/10.1103/PhysRevLett.99.236809} {\bibfield  {journal}
  {\bibinfo  {journal} {Phys. Rev. Lett.}\ }\textbf {\bibinfo {volume} {99}},\
  \bibinfo {pages} {1} (\bibinfo {year} {2007})}\BibitemShut {NoStop}%
\bibitem [{\citenamefont {Schaibley}\ \emph {et~al.}(2016)\citenamefont
  {Schaibley}, \citenamefont {Yu}, \citenamefont {Clark}, \citenamefont
  {Rivera}, \citenamefont {Ross}, \citenamefont {Seyler}, \citenamefont {Yao},\
  and\ \citenamefont {Xu}}]{Schaibley2016}%
  \BibitemOpen
  \bibfield  {author} {\bibinfo {author} {\bibfnamefont {J.~R.}\ \bibnamefont
  {Schaibley}}, \bibinfo {author} {\bibfnamefont {H.}~\bibnamefont {Yu}},
  \bibinfo {author} {\bibfnamefont {G.}~\bibnamefont {Clark}}, \bibinfo
  {author} {\bibfnamefont {P.}~\bibnamefont {Rivera}}, \bibinfo {author}
  {\bibfnamefont {J.~S.}\ \bibnamefont {Ross}}, \bibinfo {author}
  {\bibfnamefont {K.~L.}\ \bibnamefont {Seyler}}, \bibinfo {author}
  {\bibfnamefont {W.}~\bibnamefont {Yao}},\ and\ \bibinfo {author}
  {\bibfnamefont {X.}~\bibnamefont {Xu}},\ }\bibfield  {title} {\bibinfo
  {title} {{Valleytronics in 2D materials}},\ }\bibfield  {journal} {\bibinfo
  {journal} {Nat. Rev. Mater.}\ }\textbf {\bibinfo {volume} {1}},\ \href
  {https://doi.org/10.1038/natrevmats.2016.55} {10.1038/natrevmats.2016.55}
  (\bibinfo {year} {2016})\BibitemShut {NoStop}%
\bibitem [{\citenamefont {Ni}\ \emph {et~al.}(2018)\citenamefont {Ni},
  \citenamefont {Purtseladze}, \citenamefont {Smirnova}, \citenamefont
  {Slobozhanyuk},\ and\ \citenamefont {Al{\`{u}}}}]{Ni2018}%
  \BibitemOpen
  \bibfield  {author} {\bibinfo {author} {\bibfnamefont {X.}~\bibnamefont
  {Ni}}, \bibinfo {author} {\bibfnamefont {D.}~\bibnamefont {Purtseladze}},
  \bibinfo {author} {\bibfnamefont {D.~A.}\ \bibnamefont {Smirnova}}, \bibinfo
  {author} {\bibfnamefont {A.}~\bibnamefont {Slobozhanyuk}},\ and\ \bibinfo
  {author} {\bibfnamefont {A.}~\bibnamefont {Al{\`{u}}}},\ }\bibfield  {title}
  {\bibinfo {title} {{Spin and valley polarized one-way Klein tunneling in
  photonic topological insulators}},\ }\href
  {https://doi.org/10.1126/sciadv.aap8802} {\bibfield  {journal} {\bibinfo
  {journal} {Sci. Adv.}\ }\textbf {\bibinfo {volume} {4}},\ \bibinfo {pages}
  {eaap8802} (\bibinfo {year} {2018})}\BibitemShut {NoStop}%
\bibitem [{\citenamefont {Geim}\ and\ \citenamefont
  {Novoselov}(2007)}]{Geim2007}%
  \BibitemOpen
  \bibfield  {author} {\bibinfo {author} {\bibfnamefont {A.~K.}\ \bibnamefont
  {Geim}}\ and\ \bibinfo {author} {\bibfnamefont {K.~S.}\ \bibnamefont
  {Novoselov}},\ }\bibfield  {title} {\bibinfo {title} {{The rise of
  graphene}},\ }\href {https://doi.org/http://dx.doi.org/10.1038/nmat1849}
  {\bibfield  {journal} {\bibinfo  {journal} {Nat. Mater.}\ }\textbf {\bibinfo
  {volume} {6}},\ \bibinfo {pages} {183} (\bibinfo {year} {2007})}\BibitemShut
  {NoStop}%
\bibitem [{\citenamefont {{Castro Neto}}\ \emph {et~al.}(2009)\citenamefont
  {{Castro Neto}}, \citenamefont {Guinea}, \citenamefont {Peres}, \citenamefont
  {Novoselov},\ and\ \citenamefont {Geim}}]{CastroNeto2009}%
  \BibitemOpen
  \bibfield  {author} {\bibinfo {author} {\bibfnamefont {A.~H.}\ \bibnamefont
  {{Castro Neto}}}, \bibinfo {author} {\bibfnamefont {F.}~\bibnamefont
  {Guinea}}, \bibinfo {author} {\bibfnamefont {N.~M.~R.}\ \bibnamefont
  {Peres}}, \bibinfo {author} {\bibfnamefont {K.~S.}\ \bibnamefont
  {Novoselov}},\ and\ \bibinfo {author} {\bibfnamefont {A.~K.}\ \bibnamefont
  {Geim}},\ }\bibfield  {title} {\bibinfo {title} {{The electronic properties
  of graphene}},\ }\href {https://doi.org/10.1103/RevModPhys.81.109} {\bibfield
   {journal} {\bibinfo  {journal} {Rev. Mod. Phys.}\ }\textbf {\bibinfo
  {volume} {81}},\ \bibinfo {pages} {109} (\bibinfo {year} {2009})}\BibitemShut
  {NoStop}%
\bibitem [{\citenamefont {Dean}\ \emph {et~al.}(2010)\citenamefont {Dean},
  \citenamefont {Young}, \citenamefont {Meric}, \citenamefont {Lee},
  \citenamefont {Wang}, \citenamefont {Sorgenfrei}, \citenamefont {Watanabe},
  \citenamefont {Taniguchi}, \citenamefont {Kim}, \citenamefont {Shepard},\
  and\ \citenamefont {Others}}]{Dean2010}%
  \BibitemOpen
  \bibfield  {author} {\bibinfo {author} {\bibfnamefont {C.~R.}\ \bibnamefont
  {Dean}}, \bibinfo {author} {\bibfnamefont {A.~F.}\ \bibnamefont {Young}},
  \bibinfo {author} {\bibfnamefont {I.}~\bibnamefont {Meric}}, \bibinfo
  {author} {\bibfnamefont {C.}~\bibnamefont {Lee}}, \bibinfo {author}
  {\bibfnamefont {L.}~\bibnamefont {Wang}}, \bibinfo {author} {\bibfnamefont
  {S.}~\bibnamefont {Sorgenfrei}}, \bibinfo {author} {\bibfnamefont
  {K.}~\bibnamefont {Watanabe}}, \bibinfo {author} {\bibfnamefont
  {T.}~\bibnamefont {Taniguchi}}, \bibinfo {author} {\bibfnamefont
  {P.}~\bibnamefont {Kim}}, \bibinfo {author} {\bibfnamefont {K.~L.}\
  \bibnamefont {Shepard}},\ and\ \bibinfo {author} {\bibnamefont {Others}},\
  }\bibfield  {title} {\bibinfo {title} {{Boron nitride substrates for
  high-quality graphene electronics}},\ }\href@noop {} {\bibfield  {journal}
  {\bibinfo  {journal} {Nat. Nanotechnol.}\ }\textbf {\bibinfo {volume} {5}},\
  \bibinfo {pages} {722} (\bibinfo {year} {2010})}\BibitemShut {NoStop}%
\bibitem [{\citenamefont {Li}\ \emph {et~al.}(2010)\citenamefont {Li},
  \citenamefont {Luican}, \citenamefont {{Lopes Dos Santos}}, \citenamefont
  {{Castro Neto}}, \citenamefont {Reina}, \citenamefont {Kong},\ and\
  \citenamefont {Andrei}}]{Li2010c}%
  \BibitemOpen
  \bibfield  {author} {\bibinfo {author} {\bibfnamefont {G.}~\bibnamefont
  {Li}}, \bibinfo {author} {\bibfnamefont {A.}~\bibnamefont {Luican}}, \bibinfo
  {author} {\bibfnamefont {J.~M.~B.}\ \bibnamefont {{Lopes Dos Santos}}},
  \bibinfo {author} {\bibfnamefont {A.~H.}\ \bibnamefont {{Castro Neto}}},
  \bibinfo {author} {\bibfnamefont {A.}~\bibnamefont {Reina}}, \bibinfo
  {author} {\bibfnamefont {J.}~\bibnamefont {Kong}},\ and\ \bibinfo {author}
  {\bibfnamefont {E.~Y.}\ \bibnamefont {Andrei}},\ }\bibfield  {title}
  {\bibinfo {title} {{Observation of Van Hove singularities in twisted graphene
  layers}},\ }\href {https://doi.org/10.1038/nphys1463} {\bibfield  {journal}
  {\bibinfo  {journal} {Nat. Phys.}\ }\textbf {\bibinfo {volume} {6}},\
  \bibinfo {pages} {109} (\bibinfo {year} {2010})},\ \Eprint
  {https://arxiv.org/abs/0912.2102} {0912.2102} \BibitemShut {NoStop}%
\bibitem [{\citenamefont {{Su{\'{a}}rez Morell}}\ \emph
  {et~al.}(2010)\citenamefont {{Su{\'{a}}rez Morell}}, \citenamefont {Correa},
  \citenamefont {Vargas}, \citenamefont {Pacheco},\ and\ \citenamefont
  {Barticevic}}]{SuarezMorell2010}%
  \BibitemOpen
  \bibfield  {author} {\bibinfo {author} {\bibfnamefont {E.}~\bibnamefont
  {{Su{\'{a}}rez Morell}}}, \bibinfo {author} {\bibfnamefont {J.~D.}\
  \bibnamefont {Correa}}, \bibinfo {author} {\bibfnamefont {P.}~\bibnamefont
  {Vargas}}, \bibinfo {author} {\bibfnamefont {M.}~\bibnamefont {Pacheco}},\
  and\ \bibinfo {author} {\bibfnamefont {Z.}~\bibnamefont {Barticevic}},\
  }\bibfield  {title} {\bibinfo {title} {{Flat bands in slightly twisted
  bilayer graphene: Tight-binding calculations}},\ }\href
  {https://doi.org/10.1103/PhysRevB.82.121407} {\bibfield  {journal} {\bibinfo
  {journal} {Phys. Rev. B - Condens. Matter Mater. Phys.}\ }\textbf {\bibinfo
  {volume} {82}},\ \bibinfo {pages} {1} (\bibinfo {year} {2010})},\ \Eprint
  {https://arxiv.org/abs/1012.4320} {1012.4320} \BibitemShut {NoStop}%
\bibitem [{\citenamefont {Bistritzer}\ and\ \citenamefont
  {MacDonald}(2011)}]{Bistritzer2011}%
  \BibitemOpen
  \bibfield  {author} {\bibinfo {author} {\bibfnamefont {R.}~\bibnamefont
  {Bistritzer}}\ and\ \bibinfo {author} {\bibfnamefont {A.~H.}\ \bibnamefont
  {MacDonald}},\ }\bibfield  {title} {\bibinfo {title} {{Moire bands in twisted
  double-layer graphene.}},\ }\href {https://doi.org/10.1073/pnas.1108174108}
  {\bibfield  {journal} {\bibinfo  {journal} {Proc. Natl. Acad. Sci. U. S. A.}\
  }\textbf {\bibinfo {volume} {108}},\ \bibinfo {pages} {12233} (\bibinfo
  {year} {2011})},\ \Eprint {https://arxiv.org/abs/1009.4203} {1009.4203}
  \BibitemShut {NoStop}%
\bibitem [{\citenamefont {{Trambly De Laissardi{\`{e}}re}}\ \emph
  {et~al.}(2012)\citenamefont {{Trambly De Laissardi{\`{e}}re}}, \citenamefont
  {Mayou},\ and\ \citenamefont {Magaud}}]{TramblyDeLaissardiere2012}%
  \BibitemOpen
  \bibfield  {author} {\bibinfo {author} {\bibfnamefont {G.}~\bibnamefont
  {{Trambly De Laissardi{\`{e}}re}}}, \bibinfo {author} {\bibfnamefont
  {D.}~\bibnamefont {Mayou}},\ and\ \bibinfo {author} {\bibfnamefont
  {L.}~\bibnamefont {Magaud}},\ }\bibfield  {title} {\bibinfo {title}
  {{Numerical studies of confined states in rotated bilayers of graphene}},\
  }\href {https://doi.org/10.1103/PhysRevB.86.125413} {\bibfield  {journal}
  {\bibinfo  {journal} {Phys. Rev. B}\ }\textbf {\bibinfo {volume} {86}},\
  \bibinfo {pages} {1} (\bibinfo {year} {2012})},\ \Eprint
  {https://arxiv.org/abs/1203.3144} {arXiv:1203.3144} \BibitemShut {NoStop}%
\bibitem [{\citenamefont {Weckbecker}\ \emph {et~al.}(2016)\citenamefont
  {Weckbecker}, \citenamefont {Shallcross}, \citenamefont {Fleischmann},
  \citenamefont {Ray}, \citenamefont {Sharma},\ and\ \citenamefont
  {Pankratov}}]{Weckbecker2016}%
  \BibitemOpen
  \bibfield  {author} {\bibinfo {author} {\bibfnamefont {D.}~\bibnamefont
  {Weckbecker}}, \bibinfo {author} {\bibfnamefont {S.}~\bibnamefont
  {Shallcross}}, \bibinfo {author} {\bibfnamefont {M.}~\bibnamefont
  {Fleischmann}}, \bibinfo {author} {\bibfnamefont {N.}~\bibnamefont {Ray}},
  \bibinfo {author} {\bibfnamefont {S.}~\bibnamefont {Sharma}},\ and\ \bibinfo
  {author} {\bibfnamefont {O.}~\bibnamefont {Pankratov}},\ }\bibfield  {title}
  {\bibinfo {title} {{Low-energy theory for the graphene twist bilayer}},\
  }\href {https://doi.org/10.1103/PhysRevB.93.035452} {\bibfield  {journal}
  {\bibinfo  {journal} {Phys. Rev. B}\ }\textbf {\bibinfo {volume} {93}},\
  \bibinfo {pages} {1} (\bibinfo {year} {2016})}\BibitemShut {NoStop}%
\bibitem [{\citenamefont {Zhang}\ \emph {et~al.}(2013)\citenamefont {Zhang},
  \citenamefont {MacDonald},\ and\ \citenamefont {Mele}}]{Zhang2013}%
  \BibitemOpen
  \bibfield  {author} {\bibinfo {author} {\bibfnamefont {F.}~\bibnamefont
  {Zhang}}, \bibinfo {author} {\bibfnamefont {A.~H.}\ \bibnamefont
  {MacDonald}},\ and\ \bibinfo {author} {\bibfnamefont {E.~J.}\ \bibnamefont
  {Mele}},\ }\bibfield  {title} {\bibinfo {title} {{Valley Chern Numbers and
  Boundary Modes in Gapped Bilayer Graphene}}\ }\href
  {https://doi.org/10.1073/pnas.1308853110} {10.1073/pnas.1308853110} (\bibinfo
  {year} {2013}),\ \Eprint {https://arxiv.org/abs/1301.4205} {arXiv:1301.4205}
  \BibitemShut {NoStop}%
\bibitem [{\citenamefont {Yin}\ \emph {et~al.}(2016)\citenamefont {Yin},
  \citenamefont {Jiang}, \citenamefont {Qiao},\ and\ \citenamefont
  {He}}]{Yin2016}%
  \BibitemOpen
  \bibfield  {author} {\bibinfo {author} {\bibfnamefont {L.~J.}\ \bibnamefont
  {Yin}}, \bibinfo {author} {\bibfnamefont {H.}~\bibnamefont {Jiang}}, \bibinfo
  {author} {\bibfnamefont {J.~B.}\ \bibnamefont {Qiao}},\ and\ \bibinfo
  {author} {\bibfnamefont {L.}~\bibnamefont {He}},\ }\bibfield  {title}
  {\bibinfo {title} {{Direct imaging of topological edge states at a bilayer
  graphene domain wall}},\ }\href {https://doi.org/10.1038/ncomms11760}
  {\bibfield  {journal} {\bibinfo  {journal} {Nat. Commun.}\ }\textbf {\bibinfo
  {volume} {7}},\ \bibinfo {pages} {1} (\bibinfo {year} {2016})},\ \Eprint
  {https://arxiv.org/abs/1511.06498} {arXiv:1511.06498} \BibitemShut {NoStop}%
\bibitem [{\citenamefont {San-Jose}\ \emph {et~al.}(2014)\citenamefont
  {San-Jose}, \citenamefont {Guti{\'{e}}rrez-Rubio}, \citenamefont {Sturla},\
  and\ \citenamefont {Guinea}}]{SanJose2014a}%
  \BibitemOpen
  \bibfield  {author} {\bibinfo {author} {\bibfnamefont {P.}~\bibnamefont
  {San-Jose}}, \bibinfo {author} {\bibfnamefont {A.}~\bibnamefont
  {Guti{\'{e}}rrez-Rubio}}, \bibinfo {author} {\bibfnamefont {M.}~\bibnamefont
  {Sturla}},\ and\ \bibinfo {author} {\bibfnamefont {F.}~\bibnamefont
  {Guinea}},\ }\bibfield  {title} {\bibinfo {title} {{Electronic structure of
  spontaneously strained graphene on hexagonal boron nitride}},\ }\href
  {https://doi.org/10.1103/PhysRevB.90.115152} {\bibfield  {journal} {\bibinfo
  {journal} {Phys. Rev. B - Condens. Matter Mater. Phys.}\ }\textbf {\bibinfo
  {volume} {90}},\ \bibinfo {pages} {1} (\bibinfo {year} {2014})},\ \Eprint
  {https://arxiv.org/abs/1406.5999} {arXiv:1406.5999} \BibitemShut {NoStop}%
\bibitem [{\citenamefont {Huang}\ \emph {et~al.}(2018)\citenamefont {Huang},
  \citenamefont {Kim}, \citenamefont {Efimkin}, \citenamefont {Lovorn},
  \citenamefont {Taniguchi}, \citenamefont {Watanabe}, \citenamefont
  {MacDonald}, \citenamefont {Tutuc},\ and\ \citenamefont {LeRoy}}]{Huang2018}%
  \BibitemOpen
  \bibfield  {author} {\bibinfo {author} {\bibfnamefont {S.}~\bibnamefont
  {Huang}}, \bibinfo {author} {\bibfnamefont {K.}~\bibnamefont {Kim}}, \bibinfo
  {author} {\bibfnamefont {D.~K.}\ \bibnamefont {Efimkin}}, \bibinfo {author}
  {\bibfnamefont {T.}~\bibnamefont {Lovorn}}, \bibinfo {author} {\bibfnamefont
  {T.}~\bibnamefont {Taniguchi}}, \bibinfo {author} {\bibfnamefont
  {K.}~\bibnamefont {Watanabe}}, \bibinfo {author} {\bibfnamefont {A.~H.}\
  \bibnamefont {MacDonald}}, \bibinfo {author} {\bibfnamefont {E.}~\bibnamefont
  {Tutuc}},\ and\ \bibinfo {author} {\bibfnamefont {B.~J.}\ \bibnamefont
  {LeRoy}},\ }\bibfield  {title} {\bibinfo {title} {{Emergence of Topologically
  Protected Helical States in Minimally Twisted Bilayer Graphene}},\ }\href
  {https://doi.org/10.1103/PhysRevLett.121.037702} {\bibfield  {journal}
  {\bibinfo  {journal} {Phys. Rev. Lett.}\ }\textbf {\bibinfo {volume} {121}},\
  \bibinfo {pages} {37702} (\bibinfo {year} {2018})},\ \Eprint
  {https://arxiv.org/abs/1802.02999} {arXiv:1802.02999} \BibitemShut {NoStop}%
\bibitem [{\citenamefont {Cao}\ \emph {et~al.}(2018)\citenamefont {Cao},
  \citenamefont {Fatemi}, \citenamefont {Fang}, \citenamefont {Watanabe},
  \citenamefont {Taniguchi}, \citenamefont {Kaxiras},\ and\ \citenamefont
  {Jarillo-Herrero}}]{Cao2018}%
  \BibitemOpen
  \bibfield  {author} {\bibinfo {author} {\bibfnamefont {Y.}~\bibnamefont
  {Cao}}, \bibinfo {author} {\bibfnamefont {V.}~\bibnamefont {Fatemi}},
  \bibinfo {author} {\bibfnamefont {S.}~\bibnamefont {Fang}}, \bibinfo {author}
  {\bibfnamefont {K.}~\bibnamefont {Watanabe}}, \bibinfo {author}
  {\bibfnamefont {T.}~\bibnamefont {Taniguchi}}, \bibinfo {author}
  {\bibfnamefont {E.}~\bibnamefont {Kaxiras}},\ and\ \bibinfo {author}
  {\bibfnamefont {P.}~\bibnamefont {Jarillo-Herrero}},\ }\bibfield  {title}
  {\bibinfo {title} {Unconventional superconductivity in magic-angle graphene
  superlattices},\ }\href@noop {} {\bibfield  {journal} {\bibinfo  {journal}
  {Nature}\ }\textbf {\bibinfo {volume} {556}},\ \bibinfo {pages} {43}
  (\bibinfo {year} {2018})}\BibitemShut {NoStop}%
\bibitem [{\citenamefont {Po}\ \emph {et~al.}(2018)\citenamefont {Po},
  \citenamefont {Zou}, \citenamefont {Vishwanath},\ and\ \citenamefont
  {Senthil}}]{Po2018}%
  \BibitemOpen
  \bibfield  {author} {\bibinfo {author} {\bibfnamefont {H.~C.}\ \bibnamefont
  {Po}}, \bibinfo {author} {\bibfnamefont {L.}~\bibnamefont {Zou}}, \bibinfo
  {author} {\bibfnamefont {A.}~\bibnamefont {Vishwanath}},\ and\ \bibinfo
  {author} {\bibfnamefont {T.}~\bibnamefont {Senthil}},\ }\bibfield  {title}
  {\bibinfo {title} {Origin of mott insulating behavior and superconductivity
  in twisted bilayer graphene},\ }\href@noop {} {\bibfield  {journal} {\bibinfo
   {journal} {arXiv}\ } (\bibinfo {year} {2018})},\ \Eprint
  {https://arxiv.org/abs/1803.09742} {arXiv:1803.09742} \BibitemShut {NoStop}%
\bibitem [{\citenamefont {Roy}\ and\ \citenamefont {Juricic}(2018)}]{Roy2018}%
  \BibitemOpen
  \bibfield  {author} {\bibinfo {author} {\bibfnamefont {B.}~\bibnamefont
  {Roy}}\ and\ \bibinfo {author} {\bibfnamefont {V.}~\bibnamefont {Juricic}},\
  }\bibfield  {title} {\bibinfo {title} {Unconventional superconductivity in
  nearly flat bands in twisted bilayer graphene},\ }\href@noop {} {\bibfield
  {journal} {\bibinfo  {journal} {arXiv}\ } (\bibinfo {year} {2018})},\ \Eprint
  {https://arxiv.org/abs/1803.11190} {arXiv:1803.11190} \BibitemShut {NoStop}%
\bibitem [{\citenamefont {Zhou}\ \emph {et~al.}(2007)\citenamefont {Zhou},
  \citenamefont {Gweon}, \citenamefont {Fedorov}, \citenamefont {First},
  \citenamefont {de~Heer}, \citenamefont {Lee}, \citenamefont {Guinea},
  \citenamefont {{Castro Neto}},\ and\ \citenamefont {Lanzara}}]{Zhou2007}%
  \BibitemOpen
  \bibfield  {author} {\bibinfo {author} {\bibfnamefont {S.~Y.}\ \bibnamefont
  {Zhou}}, \bibinfo {author} {\bibfnamefont {G.-H.}\ \bibnamefont {Gweon}},
  \bibinfo {author} {\bibfnamefont {A.~V.}\ \bibnamefont {Fedorov}}, \bibinfo
  {author} {\bibfnamefont {P.~N.}\ \bibnamefont {First}}, \bibinfo {author}
  {\bibfnamefont {W.~A.}\ \bibnamefont {de~Heer}}, \bibinfo {author}
  {\bibfnamefont {D.-H.}\ \bibnamefont {Lee}}, \bibinfo {author} {\bibfnamefont
  {F.}~\bibnamefont {Guinea}}, \bibinfo {author} {\bibfnamefont {A.~H.}\
  \bibnamefont {{Castro Neto}}},\ and\ \bibinfo {author} {\bibfnamefont
  {A.}~\bibnamefont {Lanzara}},\ }\bibfield  {title} {\bibinfo {title}
  {{Substrate-induced bandgap opening in epitaxial graphene.}},\ }\href
  {https://doi.org/10.1038/nmat2056} {\bibfield  {journal} {\bibinfo  {journal}
  {Nat. Mater.}\ }\textbf {\bibinfo {volume} {6}},\ \bibinfo {pages} {916}
  (\bibinfo {year} {2007})}\BibitemShut {NoStop}%
\bibitem [{\citenamefont {Giovannetti}\ \emph {et~al.}(2007)\citenamefont
  {Giovannetti}, \citenamefont {Khomyakov}, \citenamefont {Brocks},
  \citenamefont {Kelly},\ and\ \citenamefont {{Van Den
  Brink}}}]{Giovannetti2007}%
  \BibitemOpen
  \bibfield  {author} {\bibinfo {author} {\bibfnamefont {G.}~\bibnamefont
  {Giovannetti}}, \bibinfo {author} {\bibfnamefont {P.~A.}\ \bibnamefont
  {Khomyakov}}, \bibinfo {author} {\bibfnamefont {G.}~\bibnamefont {Brocks}},
  \bibinfo {author} {\bibfnamefont {P.~J.}\ \bibnamefont {Kelly}},\ and\
  \bibinfo {author} {\bibfnamefont {J.}~\bibnamefont {{Van Den Brink}}},\
  }\bibfield  {title} {\bibinfo {title} {{Substrate-induced band gap in
  graphene on hexagonal boron nitride: Ab initio density functional
  calculations}},\ }\href {https://doi.org/10.1103/PhysRevB.76.073103}
  {\bibfield  {journal} {\bibinfo  {journal} {Phys. Rev. B - Condens. Matter
  Mater. Phys.}\ }\textbf {\bibinfo {volume} {76}},\ \bibinfo {pages} {2}
  (\bibinfo {year} {2007})}\BibitemShut {NoStop}%
\bibitem [{\citenamefont {Wallbank}\ \emph {et~al.}(2013)\citenamefont
  {Wallbank}, \citenamefont {Patel}, \citenamefont {Mucha-Kruczy{\'{n}}ski},
  \citenamefont {Geim},\ and\ \citenamefont {Fal'ko}}]{Wallbank2013}%
  \BibitemOpen
  \bibfield  {author} {\bibinfo {author} {\bibfnamefont {J.~R.~R.}\
  \bibnamefont {Wallbank}}, \bibinfo {author} {\bibfnamefont {A.~A.~A.}\
  \bibnamefont {Patel}}, \bibinfo {author} {\bibfnamefont {M.}~\bibnamefont
  {Mucha-Kruczy{\'{n}}ski}}, \bibinfo {author} {\bibfnamefont {A.~K.~K.}\
  \bibnamefont {Geim}},\ and\ \bibinfo {author} {\bibfnamefont {V.~I.~I.}\
  \bibnamefont {Fal'ko}},\ }\bibfield  {title} {\bibinfo {title} {{Generic
  miniband structure of graphene on a hexagonal substrate}},\ }\href
  {https://doi.org/10.1103/PhysRevB.87.245408} {\bibfield  {journal} {\bibinfo
  {journal} {Phys. Rev. B}\ }\textbf {\bibinfo {volume} {87}},\ \bibinfo
  {pages} {245408} (\bibinfo {year} {2013})}\BibitemShut {NoStop}%
\bibitem [{\citenamefont {Moon}\ and\ \citenamefont
  {Koshino}(2014)}]{Moon2014}%
  \BibitemOpen
  \bibfield  {author} {\bibinfo {author} {\bibfnamefont {P.}~\bibnamefont
  {Moon}}\ and\ \bibinfo {author} {\bibfnamefont {M.}~\bibnamefont {Koshino}},\
  }\bibfield  {title} {\bibinfo {title} {{Electronic properties of graphene
  hexagonal boron nitride moir{\'{e}} superlattice}},\ }\href
  {https://doi.org/10.1103/PhysRevB.90.155406} {\bibfield  {journal} {\bibinfo
  {journal} {Phys. Rev. B}\ }\textbf {\bibinfo {volume} {90}},\ \bibinfo
  {pages} {155406} (\bibinfo {year} {2014})},\ \Eprint
  {https://arxiv.org/abs/1406.0668} {1406.0668} \BibitemShut {NoStop}%
\bibitem [{\citenamefont {Wallbank}\ \emph {et~al.}(2015)\citenamefont
  {Wallbank}, \citenamefont {Mucha-Kruczynski}, \citenamefont {Chen},\ and\
  \citenamefont {Fal'Ko}}]{Wallbank2015}%
  \BibitemOpen
  \bibfield  {author} {\bibinfo {author} {\bibfnamefont {J.~R.}\ \bibnamefont
  {Wallbank}}, \bibinfo {author} {\bibfnamefont {M.}~\bibnamefont
  {Mucha-Kruczynski}}, \bibinfo {author} {\bibfnamefont {X.}~\bibnamefont
  {Chen}},\ and\ \bibinfo {author} {\bibfnamefont {V.~I.}\ \bibnamefont
  {Fal'Ko}},\ }\bibfield  {title} {\bibinfo {title} {{Moir{\'{e}} superlattice
  effects in graphene/boron-nitride van der Waals heterostructures}},\ }\href
  {https://doi.org/10.1002/andp.201400204} {\bibfield  {journal} {\bibinfo
  {journal} {Ann. Phys.}\ }\textbf {\bibinfo {volume} {527}},\ \bibinfo {pages}
  {359} (\bibinfo {year} {2015})}\BibitemShut {NoStop}%
\bibitem [{\citenamefont {Song}\ \emph {et~al.}(2015)\citenamefont {Song},
  \citenamefont {Samutpraphoot},\ and\ \citenamefont {Levitov}}]{Song2015}%
  \BibitemOpen
  \bibfield  {author} {\bibinfo {author} {\bibfnamefont {J.~C.~W.}\
  \bibnamefont {Song}}, \bibinfo {author} {\bibfnamefont {P.}~\bibnamefont
  {Samutpraphoot}},\ and\ \bibinfo {author} {\bibfnamefont {L.~S.}\
  \bibnamefont {Levitov}},\ }\bibfield  {title} {\bibinfo {title} {{Topological
  Bloch bands in graphene superlattices}},\ }\href
  {https://doi.org/10.1073/pnas.1424760112} {\bibfield  {journal} {\bibinfo
  {journal} {Proc. Natl. Acad. Sci.}\ }\textbf {\bibinfo {volume} {112}},\
  \bibinfo {pages} {10879} (\bibinfo {year} {2015})}\BibitemShut {NoStop}%
\bibitem [{\citenamefont {Brown}\ \emph {et~al.}(2018)\citenamefont {Brown},
  \citenamefont {Walet},\ and\ \citenamefont {Guinea}}]{Brown2017}%
  \BibitemOpen
  \bibfield  {author} {\bibinfo {author} {\bibfnamefont {R.}~\bibnamefont
  {Brown}}, \bibinfo {author} {\bibfnamefont {N.~R.}\ \bibnamefont {Walet}},\
  and\ \bibinfo {author} {\bibfnamefont {F.}~\bibnamefont {Guinea}},\
  }\bibfield  {title} {\bibinfo {title} {{Edge Modes and Nonlocal Conductance
  in Graphene Superlattices}},\ }\href
  {https://doi.org/10.1103/PhysRevLett.120.026802} {\bibfield  {journal}
  {\bibinfo  {journal} {Phys. Rev. Lett.}\ }\textbf {\bibinfo {volume} {120}},\
  \bibinfo {pages} {1} (\bibinfo {year} {2018})},\ \Eprint
  {https://arxiv.org/abs/1707.01043} {arXiv:1707.01043} \BibitemShut {NoStop}%
\bibitem [{\citenamefont {Liu}\ \emph {et~al.}(2011)\citenamefont {Liu},
  \citenamefont {Lai}, \citenamefont {Huang},\ and\ \citenamefont
  {Chan}}]{Liu2011}%
  \BibitemOpen
  \bibfield  {author} {\bibinfo {author} {\bibfnamefont {F.}~\bibnamefont
  {Liu}}, \bibinfo {author} {\bibfnamefont {Y.}~\bibnamefont {Lai}}, \bibinfo
  {author} {\bibfnamefont {X.}~\bibnamefont {Huang}},\ and\ \bibinfo {author}
  {\bibfnamefont {C.~T.}\ \bibnamefont {Chan}},\ }\bibfield  {title} {\bibinfo
  {title} {{Dirac cones at k - =0 in phononic crystals}},\ }\href
  {https://doi.org/10.1103/PhysRevB.84.224113} {\bibfield  {journal} {\bibinfo
  {journal} {Phys. Rev. B}\ }\textbf {\bibinfo {volume} {84}},\ \bibinfo
  {pages} {1} (\bibinfo {year} {2011})}\BibitemShut {NoStop}%
\bibitem [{\citenamefont {Huang}\ \emph {et~al.}(2011)\citenamefont {Huang},
  \citenamefont {Lai}, \citenamefont {Hang}, \citenamefont {Zheng},\ and\
  \citenamefont {Chan}}]{Huang2011}%
  \BibitemOpen
  \bibfield  {author} {\bibinfo {author} {\bibfnamefont {X.}~\bibnamefont
  {Huang}}, \bibinfo {author} {\bibfnamefont {Y.}~\bibnamefont {Lai}}, \bibinfo
  {author} {\bibfnamefont {Z.~H.}\ \bibnamefont {Hang}}, \bibinfo {author}
  {\bibfnamefont {H.}~\bibnamefont {Zheng}},\ and\ \bibinfo {author}
  {\bibfnamefont {C.~T.}\ \bibnamefont {Chan}},\ }\bibfield  {title} {\bibinfo
  {title} {{Dirac cones induced by accidental degeneracy in photonic crystals
  and zero-refractive-index materials}},\ }\href
  {https://doi.org/10.1038/nmat3030} {\bibfield  {journal} {\bibinfo  {journal}
  {Nat. Mater.}\ }\textbf {\bibinfo {volume} {10}},\ \bibinfo {pages} {582}
  (\bibinfo {year} {2011})}\BibitemShut {NoStop}%
\bibitem [{\citenamefont {Sakoda}(2012)}]{Sakoda2012a}%
  \BibitemOpen
  \bibfield  {author} {\bibinfo {author} {\bibfnamefont {K.}~\bibnamefont
  {Sakoda}},\ }\bibfield  {title} {\bibinfo {title} {{Double Dirac cones in
  triangular-lattice metamaterials}},\ }\href
  {https://doi.org/10.1364/OE.20.009925} {\bibfield  {journal} {\bibinfo
  {journal} {Opt. Express}\ }\textbf {\bibinfo {volume} {20}},\ \bibinfo
  {pages} {9925} (\bibinfo {year} {2012})}\BibitemShut {NoStop}%
\bibitem [{\citenamefont {Liu}\ \emph {et~al.}(2012)\citenamefont {Liu},
  \citenamefont {Huang},\ and\ \citenamefont {Chan}}]{Liu2012}%
  \BibitemOpen
  \bibfield  {author} {\bibinfo {author} {\bibfnamefont {F.}~\bibnamefont
  {Liu}}, \bibinfo {author} {\bibfnamefont {X.}~\bibnamefont {Huang}},\ and\
  \bibinfo {author} {\bibfnamefont {C.~T.}\ \bibnamefont {Chan}},\ }\bibfield
  {title} {\bibinfo {title} {{Dirac cones at k = 0 in acoustic crystals and
  zero refractive index acoustic materials}},\ }\bibfield  {journal} {\bibinfo
  {journal} {Appl. Phys. Lett.}\ }\textbf {\bibinfo {volume} {100}},\ \href
  {https://doi.org/10.1063/1.3686907} {10.1063/1.3686907} (\bibinfo {year}
  {2012})\BibitemShut {NoStop}%
\bibitem [{\citenamefont {Chen}\ \emph {et~al.}(2014)\citenamefont {Chen},
  \citenamefont {Ni}, \citenamefont {Wu}, \citenamefont {He}, \citenamefont
  {Sun}, \citenamefont {Zheng}, \citenamefont {Lu},\ and\ \citenamefont
  {Chen}}]{Chen2014}%
  \BibitemOpen
  \bibfield  {author} {\bibinfo {author} {\bibfnamefont {Z.-G.}\ \bibnamefont
  {Chen}}, \bibinfo {author} {\bibfnamefont {X.}~\bibnamefont {Ni}}, \bibinfo
  {author} {\bibfnamefont {Y.}~\bibnamefont {Wu}}, \bibinfo {author}
  {\bibfnamefont {C.}~\bibnamefont {He}}, \bibinfo {author} {\bibfnamefont
  {X.-c.}\ \bibnamefont {Sun}}, \bibinfo {author} {\bibfnamefont {L.-y.}\
  \bibnamefont {Zheng}}, \bibinfo {author} {\bibfnamefont {M.-h.}\ \bibnamefont
  {Lu}},\ and\ \bibinfo {author} {\bibfnamefont {Y.-f.}\ \bibnamefont {Chen}},\
  }\bibfield  {title} {\bibinfo {title} {{Accidental degeneracy of double Dirac
  cones in a phononic crystal}},\ }\href {https://doi.org/s10.1038/srep04613}
  {\bibfield  {journal} {\bibinfo  {journal} {Sci. Rep.}\ }\textbf {\bibinfo
  {volume} {4}},\ \bibinfo {pages} {1} (\bibinfo {year} {2014})}\BibitemShut
  {NoStop}%
\bibitem [{\citenamefont {Bradlyn}\ \emph {et~al.}(2016)\citenamefont
  {Bradlyn}, \citenamefont {Cano}, \citenamefont {Wang}, \citenamefont
  {Vergniory}, \citenamefont {Felser}, \citenamefont {Cava},\ and\
  \citenamefont {Bernevig}}]{Bradlyn2016}%
  \BibitemOpen
  \bibfield  {author} {\bibinfo {author} {\bibfnamefont {B.}~\bibnamefont
  {Bradlyn}}, \bibinfo {author} {\bibfnamefont {J.}~\bibnamefont {Cano}},
  \bibinfo {author} {\bibfnamefont {Z.}~\bibnamefont {Wang}}, \bibinfo {author}
  {\bibfnamefont {M.~G.}\ \bibnamefont {Vergniory}}, \bibinfo {author}
  {\bibfnamefont {C.}~\bibnamefont {Felser}}, \bibinfo {author} {\bibfnamefont
  {R.~J.}\ \bibnamefont {Cava}},\ and\ \bibinfo {author} {\bibfnamefont
  {B.~A.}\ \bibnamefont {Bernevig}},\ }\bibfield  {title} {\bibinfo {title}
  {{Beyond Dirac and Weyl fermions: Unconventional quasiparticles in
  conventional crystals}},\ }\bibfield  {journal} {\bibinfo  {journal} {Science
  (80-. ).}\ }\textbf {\bibinfo {volume} {353}},\ \href
  {https://doi.org/10.1126/science.aaf5037} {10.1126/science.aaf5037} (\bibinfo
  {year} {2016}),\ \Eprint {https://arxiv.org/abs/1603.03093}
  {arXiv:1603.03093} \BibitemShut {NoStop}%
\bibitem [{\citenamefont {Xue}\ \emph {et~al.}(2011)\citenamefont {Xue},
  \citenamefont {Sanchez-Yamagishi}, \citenamefont {Bulmash}, \citenamefont
  {Jacquod}, \citenamefont {Deshpande}, \citenamefont {Watanabe}, \citenamefont
  {Taniguchi}, \citenamefont {Jarillo-Herrero},\ and\ \citenamefont
  {Leroy}}]{Xue2011}%
  \BibitemOpen
  \bibfield  {author} {\bibinfo {author} {\bibfnamefont {J.~M.}\ \bibnamefont
  {Xue}}, \bibinfo {author} {\bibfnamefont {J.}~\bibnamefont
  {Sanchez-Yamagishi}}, \bibinfo {author} {\bibfnamefont {D.}~\bibnamefont
  {Bulmash}}, \bibinfo {author} {\bibfnamefont {P.}~\bibnamefont {Jacquod}},
  \bibinfo {author} {\bibfnamefont {A.}~\bibnamefont {Deshpande}}, \bibinfo
  {author} {\bibfnamefont {K.}~\bibnamefont {Watanabe}}, \bibinfo {author}
  {\bibfnamefont {T.}~\bibnamefont {Taniguchi}}, \bibinfo {author}
  {\bibfnamefont {P.}~\bibnamefont {Jarillo-Herrero}},\ and\ \bibinfo {author}
  {\bibfnamefont {B.~J.~J.}\ \bibnamefont {Leroy}},\ }\bibfield  {title}
  {\bibinfo {title} {{Scanning tunnelling microscopy and spectroscopy of
  ultra-flat graphene on hexagonal boron nitride}},\ }\href
  {https://doi.org/10.1038/nmat2968} {\bibfield  {journal} {\bibinfo  {journal}
  {Nat. Mater.}\ }\textbf {\bibinfo {volume} {10}},\ \bibinfo {pages} {282}
  (\bibinfo {year} {2011})}\BibitemShut {NoStop}%
\bibitem [{\citenamefont {Decker}\ \emph {et~al.}(2011)\citenamefont {Decker},
  \citenamefont {Wang}, \citenamefont {Brar}, \citenamefont {Regan},
  \citenamefont {Tsai}, \citenamefont {Wu}, \citenamefont {Gannett},
  \citenamefont {Zettl},\ and\ \citenamefont {Crommie}}]{Decker2011}%
  \BibitemOpen
  \bibfield  {author} {\bibinfo {author} {\bibfnamefont {R.}~\bibnamefont
  {Decker}}, \bibinfo {author} {\bibfnamefont {Y.}~\bibnamefont {Wang}},
  \bibinfo {author} {\bibfnamefont {V.~W. V.~W.}\ \bibnamefont {Brar}},
  \bibinfo {author} {\bibfnamefont {W.}~\bibnamefont {Regan}}, \bibinfo
  {author} {\bibfnamefont {H.-Z. H.-Z.}\ \bibnamefont {Tsai}}, \bibinfo
  {author} {\bibfnamefont {Q.}~\bibnamefont {Wu}}, \bibinfo {author}
  {\bibfnamefont {W.}~\bibnamefont {Gannett}}, \bibinfo {author} {\bibfnamefont
  {A.}~\bibnamefont {Zettl}},\ and\ \bibinfo {author} {\bibfnamefont {M.~F.
  M.~F.}\ \bibnamefont {Crommie}},\ }\bibfield  {title} {\bibinfo {title}
  {{Local Electronic Properties of Graphene on a BN Substrate via Scanning
  Tunneling Microscopy}},\ }\href {https://doi.org/10.1021/nl2005115}
  {\bibfield  {journal} {\bibinfo  {journal} {Nano Lett.}\ }\textbf {\bibinfo
  {volume} {11}},\ \bibinfo {pages} {2291} (\bibinfo {year}
  {2011})}\BibitemShut {NoStop}%
\bibitem [{\citenamefont {Yankowitz}\ \emph {et~al.}(2012)\citenamefont
  {Yankowitz}, \citenamefont {Xue}, \citenamefont {Cormode}, \citenamefont
  {Sanchez-Yamagishi}, \citenamefont {Watanabe}, \citenamefont {Taniguchi},
  \citenamefont {Jarillo-Herrero}, \citenamefont {Jacquod},\ and\ \citenamefont
  {Leroy}}]{Yankowitz2012}%
  \BibitemOpen
  \bibfield  {author} {\bibinfo {author} {\bibfnamefont {M.}~\bibnamefont
  {Yankowitz}}, \bibinfo {author} {\bibfnamefont {J.}~\bibnamefont {Xue}},
  \bibinfo {author} {\bibfnamefont {D.}~\bibnamefont {Cormode}}, \bibinfo
  {author} {\bibfnamefont {J.~D. J.~D.}\ \bibnamefont {Sanchez-Yamagishi}},
  \bibinfo {author} {\bibfnamefont {K.}~\bibnamefont {Watanabe}}, \bibinfo
  {author} {\bibfnamefont {T.}~\bibnamefont {Taniguchi}}, \bibinfo {author}
  {\bibfnamefont {P.}~\bibnamefont {Jarillo-Herrero}}, \bibinfo {author}
  {\bibfnamefont {P.}~\bibnamefont {Jacquod}},\ and\ \bibinfo {author}
  {\bibfnamefont {B.~J. B.~J.}\ \bibnamefont {Leroy}},\ }\bibfield  {title}
  {\bibinfo {title} {{Emergence of superlattice Dirac points in graphene on
  hexagonal boron nitride}},\ }\href {https://doi.org/10.1038/nphys2272}
  {\bibfield  {journal} {\bibinfo  {journal} {Nat. Phys.}\ }\textbf {\bibinfo
  {volume} {8}},\ \bibinfo {pages} {382} (\bibinfo {year} {2012})}\BibitemShut
  {NoStop}%
\bibitem [{\citenamefont {Woods}\ \emph {et~al.}(2014)\citenamefont {Woods},
  \citenamefont {Britnell}, \citenamefont {Eckmann}, \citenamefont {Ma},
  \citenamefont {Lu}, \citenamefont {Guo}, \citenamefont {Lin}, \citenamefont
  {Yu}, \citenamefont {Cao}, \citenamefont {Gorbachev}, \citenamefont
  {Kretinin}, \citenamefont {Park}, \citenamefont {Ponomarenko}, \citenamefont
  {Katsnelson}, \citenamefont {Gornostyrev}, \citenamefont {Watanabe},
  \citenamefont {Taniguchi}, \citenamefont {Casiraghi}, \citenamefont {Gao},
  \citenamefont {Geim},\ and\ \citenamefont {Novoselov}}]{Woods2014}%
  \BibitemOpen
  \bibfield  {author} {\bibinfo {author} {\bibfnamefont {C.~R.}\ \bibnamefont
  {Woods}}, \bibinfo {author} {\bibfnamefont {L.}~\bibnamefont {Britnell}},
  \bibinfo {author} {\bibfnamefont {a.}~\bibnamefont {Eckmann}}, \bibinfo
  {author} {\bibfnamefont {R.~S.}\ \bibnamefont {Ma}}, \bibinfo {author}
  {\bibfnamefont {J.~C.}\ \bibnamefont {Lu}}, \bibinfo {author} {\bibfnamefont
  {H.~M.}\ \bibnamefont {Guo}}, \bibinfo {author} {\bibfnamefont
  {X.}~\bibnamefont {Lin}}, \bibinfo {author} {\bibfnamefont {G.~L.}\
  \bibnamefont {Yu}}, \bibinfo {author} {\bibfnamefont {Y.}~\bibnamefont
  {Cao}}, \bibinfo {author} {\bibfnamefont {R.~V.}\ \bibnamefont {Gorbachev}},
  \bibinfo {author} {\bibfnamefont {a.~V.}\ \bibnamefont {Kretinin}}, \bibinfo
  {author} {\bibfnamefont {J.}~\bibnamefont {Park}}, \bibinfo {author}
  {\bibfnamefont {L.~a.}\ \bibnamefont {Ponomarenko}}, \bibinfo {author}
  {\bibfnamefont {M.~I.}\ \bibnamefont {Katsnelson}}, \bibinfo {author}
  {\bibfnamefont {Y.~N.}\ \bibnamefont {Gornostyrev}}, \bibinfo {author}
  {\bibfnamefont {K.}~\bibnamefont {Watanabe}}, \bibinfo {author}
  {\bibfnamefont {T.}~\bibnamefont {Taniguchi}}, \bibinfo {author}
  {\bibfnamefont {C.}~\bibnamefont {Casiraghi}}, \bibinfo {author}
  {\bibfnamefont {H.-j.}\ \bibnamefont {Gao}}, \bibinfo {author} {\bibfnamefont
  {a.~K.}\ \bibnamefont {Geim}},\ and\ \bibinfo {author} {\bibfnamefont
  {K.~S.}\ \bibnamefont {Novoselov}},\ }\bibfield  {title} {\bibinfo {title}
  {{Commensurate–incommensurate transition in graphene on hexagonal boron
  nitride}},\ }\href {https://doi.org/10.1038/nphys2954} {\bibfield  {journal}
  {\bibinfo  {journal} {Nat. Phys.}\ }\textbf {\bibinfo {volume} {10}},\
  \bibinfo {pages} {1} (\bibinfo {year} {2014})}\BibitemShut {NoStop}%
\bibitem [{\citenamefont {Nielsen}\ and\ \citenamefont
  {Hedeg{\aa}rd}(1995)}]{Nielsen1995}%
  \BibitemOpen
  \bibfield  {author} {\bibinfo {author} {\bibfnamefont {M.}~\bibnamefont
  {Nielsen}}\ and\ \bibinfo {author} {\bibfnamefont {P.}~\bibnamefont
  {Hedeg{\aa}rd}},\ }\bibfield  {title} {\bibinfo {title} {{Two-dimensional
  electron transport in the presence of magnetic flux vortices}},\ }\href
  {https://doi.org/10.1103/PhysRevB.51.7679} {\bibfield  {journal} {\bibinfo
  {journal} {Phys. Rev. B}\ }\textbf {\bibinfo {volume} {51}},\ \bibinfo
  {pages} {7679} (\bibinfo {year} {1995})}\BibitemShut {NoStop}%
\bibitem [{\citenamefont {Moon}\ and\ \citenamefont
  {Koshino}(2012)}]{Moon2012}%
  \BibitemOpen
  \bibfield  {author} {\bibinfo {author} {\bibfnamefont {P.}~\bibnamefont
  {Moon}}\ and\ \bibinfo {author} {\bibfnamefont {M.}~\bibnamefont {Koshino}},\
  }\bibfield  {title} {\bibinfo {title} {{Energy spectrum and quantum Hall
  effect in twisted bilayer graphene}},\ }\href
  {https://doi.org/10.1103/PhysRevB.85.195458} {\bibfield  {journal} {\bibinfo
  {journal} {Phys. Rev. B}\ }\textbf {\bibinfo {volume} {85}},\ \bibinfo
  {pages} {1} (\bibinfo {year} {2012})},\ \Eprint
  {https://arxiv.org/abs/1202.4365} {arXiv:1202.4365} \BibitemShut {NoStop}%
\bibitem [{Note1()}]{Note1}%
  \BibitemOpen
  \bibinfo {note} {Including higher bands leads to a deformation of the
  planes.}\BibitemShut {Stop}%
\bibitem [{\citenamefont {Xiao}\ \emph {et~al.}(2010)\citenamefont {Xiao},
  \citenamefont {Chang},\ and\ \citenamefont {Niu}}]{Xiao2010}%
  \BibitemOpen
  \bibfield  {author} {\bibinfo {author} {\bibfnamefont {D.}~\bibnamefont
  {Xiao}}, \bibinfo {author} {\bibfnamefont {M.~C.}\ \bibnamefont {Chang}},\
  and\ \bibinfo {author} {\bibfnamefont {Q.}~\bibnamefont {Niu}},\ }\bibfield
  {title} {\bibinfo {title} {{Berry phase effects on electronic properties}},\
  }\href {https://doi.org/10.1103/RevModPhys.82.1959} {\bibfield  {journal}
  {\bibinfo  {journal} {Rev. Mod. Phys.}\ }\textbf {\bibinfo {volume} {82}},\
  \bibinfo {pages} {1959} (\bibinfo {year} {2010})}\BibitemShut {NoStop}%
\bibitem [{\citenamefont {Fukui}\ and\ \citenamefont
  {Hatsugai}(2007)}]{Fukui2007}%
  \BibitemOpen
  \bibfield  {author} {\bibinfo {author} {\bibfnamefont {T.}~\bibnamefont
  {Fukui}}\ and\ \bibinfo {author} {\bibfnamefont {Y.}~\bibnamefont
  {Hatsugai}},\ }\bibfield  {title} {\bibinfo {title} {{Quantum spin Hall
  effect in three dimensional materials: Lattice computation of Z 2 topological
  invariants and its application to Bi and Sb}},\ }\href
  {https://doi.org/10.1143/JPSJ.76.053702} {\bibfield  {journal} {\bibinfo
  {journal} {J. Phys. Soc. Japan}\ }\textbf {\bibinfo {volume} {76}},\ \bibinfo
  {pages} {1} (\bibinfo {year} {2007})}\BibitemShut {NoStop}%
\bibitem [{\citenamefont {Gorbachev}\ \emph {et~al.}(2014)\citenamefont
  {Gorbachev}, \citenamefont {Song}, \citenamefont {Yu}, \citenamefont
  {Kretinin}, \citenamefont {Withers}, \citenamefont {Cao}, \citenamefont
  {Mishchenko}, \citenamefont {Grigorieva}, \citenamefont {Novoselov},
  \citenamefont {Levitov},\ and\ \citenamefont {Geim}}]{Gorbachev2014}%
  \BibitemOpen
  \bibfield  {author} {\bibinfo {author} {\bibfnamefont {R.~V.}\ \bibnamefont
  {Gorbachev}}, \bibinfo {author} {\bibfnamefont {J.~C.}\ \bibnamefont {Song}},
  \bibinfo {author} {\bibfnamefont {G.~L.}\ \bibnamefont {Yu}}, \bibinfo
  {author} {\bibfnamefont {A.~V.}\ \bibnamefont {Kretinin}}, \bibinfo {author}
  {\bibfnamefont {F.}~\bibnamefont {Withers}}, \bibinfo {author} {\bibfnamefont
  {Y.}~\bibnamefont {Cao}}, \bibinfo {author} {\bibfnamefont {A.}~\bibnamefont
  {Mishchenko}}, \bibinfo {author} {\bibfnamefont {I.~V.}\ \bibnamefont
  {Grigorieva}}, \bibinfo {author} {\bibfnamefont {K.~S.}\ \bibnamefont
  {Novoselov}}, \bibinfo {author} {\bibfnamefont {L.~S.}\ \bibnamefont
  {Levitov}},\ and\ \bibinfo {author} {\bibfnamefont {A.~K.}\ \bibnamefont
  {Geim}},\ }\bibfield  {title} {\bibinfo {title} {{Detecting topological
  currents in graphene superlattices}},\ }\href
  {https://doi.org/10.1126/science.1254966} {\bibfield  {journal} {\bibinfo
  {journal} {Science (80-. ).}\ }\textbf {\bibinfo {volume} {346}},\ \bibinfo
  {pages} {448} (\bibinfo {year} {2014})},\ \Eprint
  {https://arxiv.org/abs/1409.0113} {arXiv:1409.0113} \BibitemShut {NoStop}%
\bibitem [{\citenamefont {Ma}\ and\ \citenamefont {Shvets}(2016)}]{Ma2016}%
  \BibitemOpen
  \bibfield  {author} {\bibinfo {author} {\bibfnamefont {T.}~\bibnamefont
  {Ma}}\ and\ \bibinfo {author} {\bibfnamefont {G.}~\bibnamefont {Shvets}},\
  }\bibfield  {title} {\bibinfo {title} {{All-Si valley-Hall photonic
  topological insulator}},\ }\bibfield  {journal} {\bibinfo  {journal} {New J.
  Phys.}\ }\textbf {\bibinfo {volume} {18}},\ \href
  {https://doi.org/10.1088/1367-2630/18/2/025012}
  {10.1088/1367-2630/18/2/025012} (\bibinfo {year} {2016})\BibitemShut
  {NoStop}%
\bibitem [{\citenamefont {Dong}\ \emph {et~al.}(2017)\citenamefont {Dong},
  \citenamefont {Chen}, \citenamefont {Zhu}, \citenamefont {Wang},\ and\
  \citenamefont {Zhang}}]{Dong2017}%
  \BibitemOpen
  \bibfield  {author} {\bibinfo {author} {\bibfnamefont {J.~W.}\ \bibnamefont
  {Dong}}, \bibinfo {author} {\bibfnamefont {X.~D.}\ \bibnamefont {Chen}},
  \bibinfo {author} {\bibfnamefont {H.}~\bibnamefont {Zhu}}, \bibinfo {author}
  {\bibfnamefont {Y.}~\bibnamefont {Wang}},\ and\ \bibinfo {author}
  {\bibfnamefont {X.}~\bibnamefont {Zhang}},\ }\bibfield  {title} {\bibinfo
  {title} {{Valley photonic crystals for control of spin and topology}},\
  }\href {https://doi.org/10.1038/nmat4807} {\bibfield  {journal} {\bibinfo
  {journal} {Nat. Mater.}\ }\textbf {\bibinfo {volume} {16}},\ \bibinfo {pages}
  {298} (\bibinfo {year} {2017})}\BibitemShut {NoStop}%
\bibitem [{\citenamefont {Kang}\ \emph {et~al.}(2018)\citenamefont {Kang},
  \citenamefont {Cheng}, \citenamefont {Ni}, \citenamefont {Khanikaev},\ and\
  \citenamefont {Genack}}]{Kang2018}%
  \BibitemOpen
  \bibfield  {author} {\bibinfo {author} {\bibfnamefont {Y.}~\bibnamefont
  {Kang}}, \bibinfo {author} {\bibfnamefont {X.}~\bibnamefont {Cheng}},
  \bibinfo {author} {\bibfnamefont {X.}~\bibnamefont {Ni}}, \bibinfo {author}
  {\bibfnamefont {A.~B.}\ \bibnamefont {Khanikaev}},\ and\ \bibinfo {author}
  {\bibfnamefont {A.~Z.}\ \bibnamefont {Genack}},\ }\bibfield  {title}
  {\bibinfo {title} {{Pseudospin-valley coupled edge states in a photonic
  topological insulator}},\ }\href {https://doi.org/10.1038/s41467-018-05408-w}
  {\bibfield  {journal} {\bibinfo  {journal} {Nat. Commun.}\ ,\ \bibinfo
  {pages} {1}} (\bibinfo {year} {2018})},\ \Eprint
  {https://arxiv.org/abs/1804.08707} {arXiv:1804.08707} \BibitemShut {NoStop}%
\bibitem [{\citenamefont {Noh}\ \emph {et~al.}(2018)\citenamefont {Noh},
  \citenamefont {Huang}, \citenamefont {Chen},\ and\ \citenamefont
  {Rechtsman}}]{Noh2018}%
  \BibitemOpen
  \bibfield  {author} {\bibinfo {author} {\bibfnamefont {J.}~\bibnamefont
  {Noh}}, \bibinfo {author} {\bibfnamefont {S.}~\bibnamefont {Huang}}, \bibinfo
  {author} {\bibfnamefont {K.~P.}\ \bibnamefont {Chen}},\ and\ \bibinfo
  {author} {\bibfnamefont {M.~C.}\ \bibnamefont {Rechtsman}},\ }\bibfield
  {title} {\bibinfo {title} {{Observation of Photonic Topological Valley Hall
  Edge States}},\ }\href {https://doi.org/10.1103/PhysRevLett.120.063902}
  {\bibfield  {journal} {\bibinfo  {journal} {Phys. Rev. Lett.}\ }\textbf
  {\bibinfo {volume} {120}},\ \bibinfo {pages} {63902} (\bibinfo {year}
  {2018})},\ \Eprint {https://arxiv.org/abs/1706.00059} {arXiv:1706.00059}
  \BibitemShut {NoStop}%
\bibitem [{\citenamefont {Park}\ \emph {et~al.}(2008)\citenamefont {Park},
  \citenamefont {Yang}, \citenamefont {Son}, \citenamefont {Cohen},\ and\
  \citenamefont {Louie}}]{Park2008a}%
  \BibitemOpen
  \bibfield  {author} {\bibinfo {author} {\bibfnamefont {C.~H.}\ \bibnamefont
  {Park}}, \bibinfo {author} {\bibfnamefont {L.}~\bibnamefont {Yang}}, \bibinfo
  {author} {\bibfnamefont {Y.~W.}\ \bibnamefont {Son}}, \bibinfo {author}
  {\bibfnamefont {M.~L.}\ \bibnamefont {Cohen}},\ and\ \bibinfo {author}
  {\bibfnamefont {S.~G.}\ \bibnamefont {Louie}},\ }\bibfield  {title} {\bibinfo
  {title} {{New generation of massless dirac fermions in graphene under
  external periodic potentials}},\ }\href
  {https://doi.org/10.1103/PhysRevLett.101.126804} {\bibfield  {journal}
  {\bibinfo  {journal} {Phys. Rev. Lett.}\ }\textbf {\bibinfo {volume} {101}},\
  \bibinfo {pages} {1} (\bibinfo {year} {2008})},\ \Eprint
  {https://arxiv.org/abs/0809.3422} {0809.3422} \BibitemShut {NoStop}%
\bibitem [{\citenamefont {Ye}\ and\ \citenamefont {Qi}(2011)}]{Ye2011}%
  \BibitemOpen
  \bibfield  {author} {\bibinfo {author} {\bibfnamefont {X.}~\bibnamefont
  {Ye}}\ and\ \bibinfo {author} {\bibfnamefont {L.}~\bibnamefont {Qi}},\
  }\bibfield  {title} {\bibinfo {title} {{Two-dimensionally patterned
  nanostructures based on monolayer colloidal crystals: Controllable
  fabrication, assembly, and applications}},\ }\href
  {https://doi.org/10.1016/j.nantod.2011.10.002} {\bibfield  {journal}
  {\bibinfo  {journal} {Nano Today}\ }\textbf {\bibinfo {volume} {6}},\
  \bibinfo {pages} {608} (\bibinfo {year} {2011})}\BibitemShut {NoStop}%
\bibitem [{\citenamefont {Jung}\ \emph {et~al.}(2015)\citenamefont {Jung},
  \citenamefont {DaSilva}, \citenamefont {MacDonald},\ and\ \citenamefont
  {Adam}}]{Jung2015}%
  \BibitemOpen
  \bibfield  {author} {\bibinfo {author} {\bibfnamefont {J.}~\bibnamefont
  {Jung}}, \bibinfo {author} {\bibfnamefont {A.~M.}\ \bibnamefont {DaSilva}},
  \bibinfo {author} {\bibfnamefont {A.~H.}\ \bibnamefont {MacDonald}},\ and\
  \bibinfo {author} {\bibfnamefont {S.}~\bibnamefont {Adam}},\ }\bibfield
  {title} {\bibinfo {title} {{Origin of band gaps in graphene on hexagonal
  boron nitride}},\ }\href {https://doi.org/10.1038/ncomms7308} {\bibfield
  {journal} {\bibinfo  {journal} {Nat. Commun.}\ }\textbf {\bibinfo {volume}
  {6}},\ \bibinfo {pages} {6308} (\bibinfo {year} {2015})},\ \Eprint
  {https://arxiv.org/abs/1403.0496} {arXiv:1403.0496} \BibitemShut {NoStop}%
\bibitem [{\citenamefont {Uehlinger}\ \emph {et~al.}(2013)\citenamefont
  {Uehlinger}, \citenamefont {Jotzu}, \citenamefont {Messer}, \citenamefont
  {Greif}, \citenamefont {Hofstetter}, \citenamefont {Bissbort},\ and\
  \citenamefont {Esslinger}}]{Uehlinger2013}%
  \BibitemOpen
  \bibfield  {author} {\bibinfo {author} {\bibfnamefont {T.}~\bibnamefont
  {Uehlinger}}, \bibinfo {author} {\bibfnamefont {G.}~\bibnamefont {Jotzu}},
  \bibinfo {author} {\bibfnamefont {M.}~\bibnamefont {Messer}}, \bibinfo
  {author} {\bibfnamefont {D.}~\bibnamefont {Greif}}, \bibinfo {author}
  {\bibfnamefont {W.}~\bibnamefont {Hofstetter}}, \bibinfo {author}
  {\bibfnamefont {U.}~\bibnamefont {Bissbort}},\ and\ \bibinfo {author}
  {\bibfnamefont {T.}~\bibnamefont {Esslinger}},\ }\bibfield  {title} {\bibinfo
  {title} {{Artificial graphene with tunable interactions}},\ }\href
  {https://doi.org/10.1103/PhysRevLett.111.185307} {\bibfield  {journal}
  {\bibinfo  {journal} {Phys. Rev. Lett.}\ }\textbf {\bibinfo {volume} {111}},\
  \bibinfo {pages} {1} (\bibinfo {year} {2013})},\ \Eprint
  {https://arxiv.org/abs/1308.4401} {1308.4401} \BibitemShut {NoStop}%
\end{thebibliography}%


\end{document}